\documentclass[twocolumn]{aastex701}
\usepackage{amsmath}
\usepackage{xspace}
\usepackage{xcolor}
\usepackage{tikz-imagelabels}
\usepackage{soul}

\title{The Extended Baryonic Tully-Fisher Relation for MaNGA Galaxies}

\newcommand{\HI}{\ion{H}{1}\xspace}
\newcommand{\Htwo}{H$_2$\xspace}
\newcommand{\MHI}{$M_\text{H{\sc i}}$\xspace}
\newcommand{\Halpha}{H$\alpha$\xspace} 

\newcommand{\Mbar}{$M_{\rm bar}$\xspace}
\newcommand{\Mtot}{$M_{\rm tot}$\xspace}

\shorttitle{MaNGA BTFR}
\shortauthors{Ravi, Douglass \& Demina}

\begin{document}

\title{The Extended Baryonic Tully-Fisher Relation for SDSS MaNGA Galaxies}

\correspondingauthor{Nitya Ravi}
\email{nravi3@ur.rochester.edu}

\author[0000-0002-4248-2840]{Nitya Ravi}
\affiliation{Department of Physics \& Astronomy, University of Rochester, 500 Joseph C. Wilson Blvd., Rochester, NY 14627} 
\email{nravi3@ur.rochester.edu}
 
\author[0000-0002-9540-546X]{Kelly A. Douglass}
\affiliation{Department of Physics \& Astronomy, University of Rochester, 500 Joseph C. Wilson Blvd., Rochester, NY 14627} 
\email{kdougla7@ur.rochester.edu}

\author[0000-0002-7852-167X]{Regina Demina}
\affiliation{Department of Physics \& Astronomy, University of Rochester, 500 Joseph C. Wilson Blvd., Rochester, NY 14627} 
\email{regina@pas.rochester.edu}


\begin{abstract}

The baryonic Tully-Fisher relation (BTFR), a relationship between rotational velocity and baryonic mass in spiral galaxies, probes the relative content of baryonic and dark matter in galaxies and thus provides a good test of $\Lambda$CDM. Using H$\alpha$ kinematics to model the rotation curves of spiral galaxies, we construct the BTFR for 5743 SDSS MaNGA DR17 galaxies. To extend the BTFR to higher masses using elliptical galaxies, we estimate their total masses from their stellar velocity dispersions using the virial theorem and define the effective rotational velocity as the velocity a rotation-supported galaxy would exhibit given this mass. The baryonic mass of spiral galaxies is composed of stellar, \HI, \Htwo, and He mass, while only the stellar mass is used for the baryonic content of ellipticals. We construct and fit the BTFR for a matched subsample of spiral and elliptical MaNGA and IllustrisTNG 100-1 (TNG100) galaxies, finding BTFR slopes between 3.2 and 4.0. We fit a joint BTFR for the 5743 MaNGA spiral and elliptical galaxies and find a BTFR slope of $3.54 \substack{+0.65 \\ -0.48}$, which is in good agreement with TNG100 galaxies with baryonic masses greater than $10^9 M_{\odot}$ for which we find a BTFR slope of $3.57\substack{+0.48 \\ -0.37}$. Within this mass range, the MaNGA galaxies are consistent with both the $\Lambda$CDM simulation and the prediction from MOND; a sample of lower mass galaxies is necessary to differentiate between the two models. 
\end{abstract}

\section{Introduction}\label{sec:intro}

The Tully-Fisher relation (TFR) is an empirical relationship between spiral galaxies' rotational velocities and their luminosities. This relationship was first reported by \cite{Tully77} as the correlation between the \HI line width and the size or absolute magnitude of spiral galaxies and has since been observed in different measures of rotational velocity and luminosity \citep[e.g.,][]{Mocz12, Ristea24}. The TFR can be used to measure distances to galaxies: if this relation is the same for all spiral galaxies, the distance to any galaxy can be calculated given its rotational velocity and apparent magnitude.

If we assume that the galaxy luminosity is a proxy for stellar mass, and the rotational velocity is a proxy for a galaxy's dynamical mass, the TFR quantifies the relationship between these two masses. However, there are indications of deviations from the TFR in spiral galaxies at both low and high mass end. For high-mass galaxies, there is a sign of a turnover: the luminosity stays the same while the rotational velocity continues to increase \citep[e.g.,][]{Kourkchi20}. For low-mass galaxies, the rotational velocities appear to be too high for their luminosities \citep[e.g.,][]{McGaugh:2000}. The low-luminosity galaxies tend to have a higher gas content compared to the higher luminosity galaxies. To help reduce the discrepancy at the low-mass end, we can construct the baryonic Tully-Fisher relation (BTFR): the relationship between spiral galaxies' rotational velocities and baryonic masses, which in addition to stellar also includes the gas mass.

The BTFR can be used to test the theory of Modified Newtonian Dynamics \citep[MOND;][]{Milgrom83}. MOND proposes a dynamic explanation for flat rotation curves in spiral galaxies, eliminating the need for dark matter. In the small acceleration limit, the gravitational acceleration deviates from Newtonian and approaches $\sqrt{a_0 g_N}$, where $g_N$ is the Newtonian acceleration and $a_0$ is calculated empirically \citep{Sanders02}. MOND predicts spiral galaxies to follow a BTFR with a slope of 4 without significant scatter \citep{McGaugh:2012ac}. In contrast, $\Lambda$CDM explains the flat rotation curves by the presence of dark matter, the amount of which can vary depending on galaxy formation history and the baryon fraction in halos. Thus, $\Lambda$CDM predicts the BTFR with the slope between 3 and 4, significant scatter, and a possible turnover at high masses \citep{Bradford16}.

We can evaluate the consistency of $\Lambda$CDM predictions by comparing the BTFR of simulated and observed galaxies. For example, \cite{Goddy23} compared 377 galaxies in the \HI-MaNGA survey to mock \HI observations of galaxies in the IllustrisTNG 100-1 (TNG100) simulation and found the BTFR to be consistent within uncertainties. In this paper, we use the final MaNGA catalog to increase the statistical significance of spiral galaxies, whose rotation curves we model with the MaNGA \Halpha emission-line velocity maps. In addition, we propose a method to extend the BTFR to the more massive elliptical galaxies, and thus probe for a potential turnover.

In contrast to spiral galaxies, which are supported by rotation, elliptical galaxies are supported against gravitational collapse by pressure. According to the virial theorem, if an elliptical galaxy is in equilibrium, its total, or gravitational, mass is related to the mean velocity of the stars, the stellar mass, and the half-mass radius, which we approximate by the half-light radius \citep{Ryden20}. From the total mass, we calculate an effective rotational velocity, which we define as the rotational velocity an elliptical galaxy would exhibit if it were rotationally supported. This strategy allows us to extend the BTFR to elliptical galaxies and examine if the linear relationship between baryonic mass and rotational velocity holds at higher masses.

The paper is organized as follows. In Sec.~\ref{sec:galaxy_selection}, we describe the data used in this analysis. In Sec.~\ref{sec:method} we introduce the methodology, which includes the evaluation of the stellar, baryonic, and total masses for the spiral and elliptical galaxies, and the definition of the effective rotational velocity for the elliptical galaxies. In Sec.~\ref{sec:verification}, we verify this methodology using the IllustrisTNG simulation. In Sec.~\ref{sec:baryonic_mass_comparison}, we discuss the effect of the galaxy's evolutionary stage on its mass-to-luminosity relationship. In Sec.~\ref{sec:btfr}, we construct the BTFR for the MaNGA and IllustrisTNG galaxies and evaluate its consistency with the $\Lambda$CDM and MOND hypotheses. We conclude in Sec.~\ref{sec:conclusion}.

\section{Data sets and Galaxy Selection} \label{sec:galaxy_selection}

\subsection{SDSS MaNGA DR17}

We use data from the SDSS MaNGA DR17 \citep{MaNGA, SDSS17} to estimate the total and stellar masses of spiral and elliptical galaxies. SDSS MaNGA was an integral field spectroscopic survey that observed over 10,000 galaxies. Spectra across a galaxy are measured with an integral field unit (IFU), a bundle of 19 to 127 spectroscopic fibers arranged in a hexagon. Each IFU covers {12.5\arcsec} to {32.5\arcsec} in diameter \citep{Law15}. The light from the IFUs is received by two spectrographs covering wavelength ranges of 3600--6000{\AA} and 6000--10300{\AA} with a resolution $\lambda/\Delta \lambda \sim 2000$ \citep{Smee13}.

We make use of each galaxy's \Halpha velocity map, $g$-band weighted mean flux map, and stellar velocity dispersion map from the MaNGA Data Analysis Pipeline \citep[DAP;][]{Westfall19}. We use the stellar mass density maps from the Pipe3D pipeline \citep{Sanchez16, Sanchez22, Lacerda22}. All distances are given in units of kpc $h^{-1}$, where $h$ is the reduced Hubble constant defined by $H_0 = 100 h$ km s$^{-1}$ Mpc$^{-1}$.

\subsection{SDSS DR7}

SDSS DR7 \citep{SDSS7} has wide-band photometry in the SDSS $u$, $g$, $r$, $i$, and $z$ bands over 10,000 deg$^2$ of the northern sky, observed with a wide-field 2.5~m telescope at the Apache Point Observatory in New Mexico. SDSS DR7, conducted with a fiber-fed pair of double spectrographs covering a wavelength range of 3800 \AA-- 9200\AA, included follow-up spectroscopy of over 90,000 galaxies. The galaxies chosen for follow-up spectroscopy include a sample complete to a Petrosian $r$-band magnitude, $m_r$, of 17.77 \citep{Strauss02} and additional samples of luminous red galaxies, resulting in a sample that is volume- (flux-) limited to a redshift of 0.38 (0.55) \citep{Eisenstein01}.

We obtain each galaxy's $u-r$ color, $\Delta (g-i)$ color gradient, and inverse concentration index, $c_{\rm inv}$ from the Korea Institute for Advanced Study Value-Added Galaxy Catalog \citep[KIAS-VAGC;][]{Choi10}, which uses the independent SDSS DR7 data reduction from the New York University Value-Added Galaxy Catalog \citep[NYU-VAGC;][]{Blanton05}. We extract the absolute magnitude, the 90\% and 50\% elliptical Petrosian radii, $R_{90}$ and $R_{50}$ respectively, and stellar mass for each galaxy from version 1.0.1 of the NASA Sloan Atlas \citep[NSA;][]{Blanton11}.

\subsection{IllustrisTNG}

IllustrisTNG is a suite of cosmological simulations using the quasi-Lagrangian code \texttt{AREPO} \citep{Nelson19, Pillepich18b, Nelson18, Naiman18, Marinacci18, Springel18, Springel10}. IllustrisTNG evolves galaxies from redshift $z=127$ to $z=0$ assuming \cite{Planck16} cosmology, with star particles defined as stellar populations with a \cite{Chabrier03} initial mass function (IMF). We use data from the TNG100-1 (TNG100) simulation which has a volume of (110.7~Mpc)$^3$. In order to compare simulation galaxies with SDSS MaNGA data, we take galaxies from the $z=0$ snapshot.

\subsection{Defining the Spiral and Elliptical Samples}

We classify galaxies as either spiral or elliptical using their kinematics and photometry. For our purposes, spiral galaxies are defined as those supported by rotation, while elliptical galaxies are supported by pressure. Rotation-supported spiral galaxies will then have a smooth \Halpha velocity map, while pressure-supported elliptical galaxies will have a randomized velocity map. We quantify the smoothness of the velocity map as described in \cite{Douglass19}. The T-type of each galaxy as classified by the MaNGA Morphology Deep Learning (DL) DR17 Value Added Catalog \citep{DominguezSanchez18, DominguezSanchez22} is also used to help classify the galaxies as either elliptical or spiral. In total, we define
\begin{description}
    \item[Spiral] T-type $> 0$ and a ``smoothness score'' $< 2.0$
    \item[Elliptical] T-type $< 0$ and a ``smoothness score'' $> 2.0$
\end{description}
This results in a sample of 2460 elliptical galaxies and 5626 spiral galaxies from SDSS MaNGA DR17.

\subsection{Color-Magnitude Diagram Classification}\label{sec:cmd}

\begin{figure}
    \centering
    \includegraphics[width=0.49\textwidth]{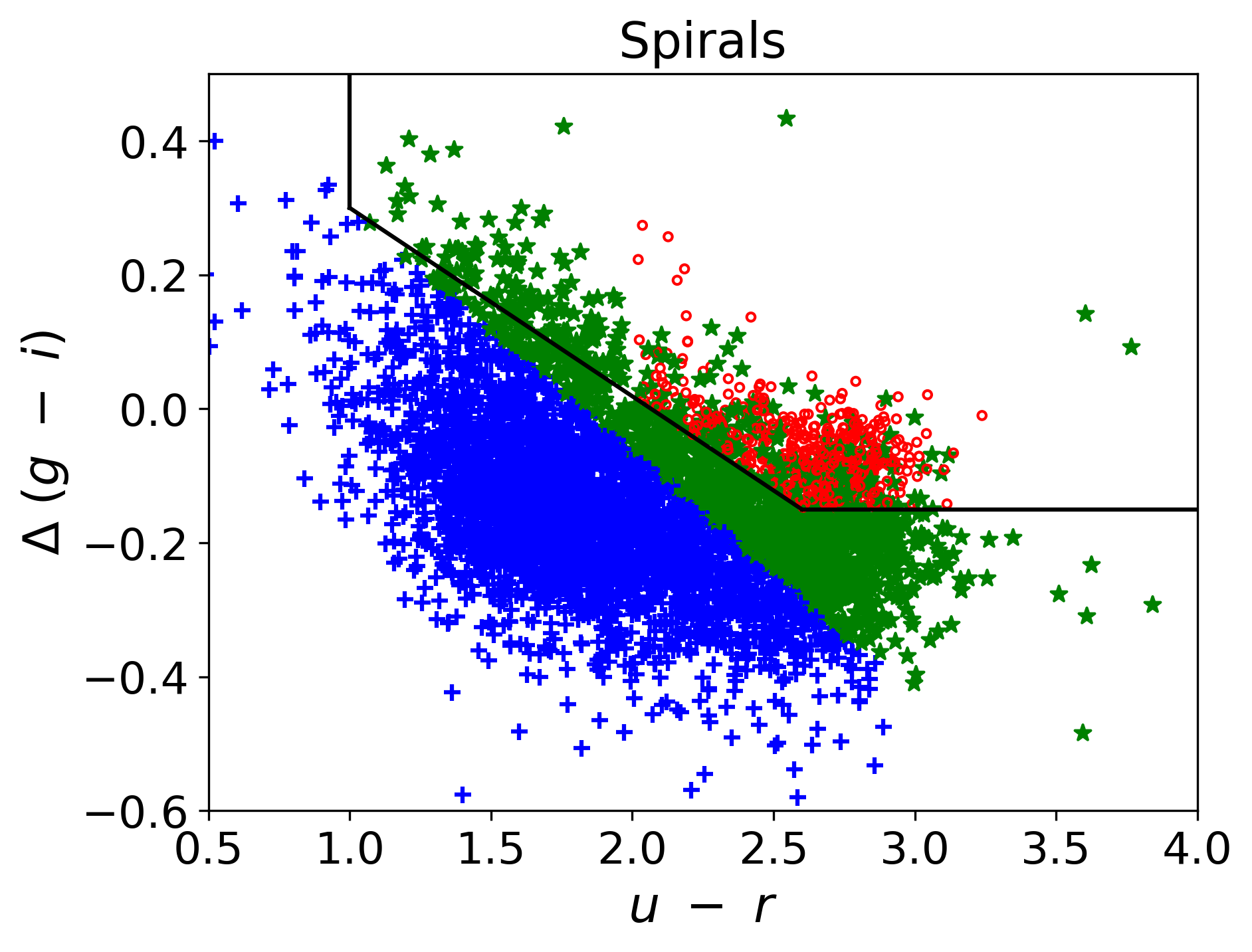}
    \includegraphics[width=0.49\textwidth]{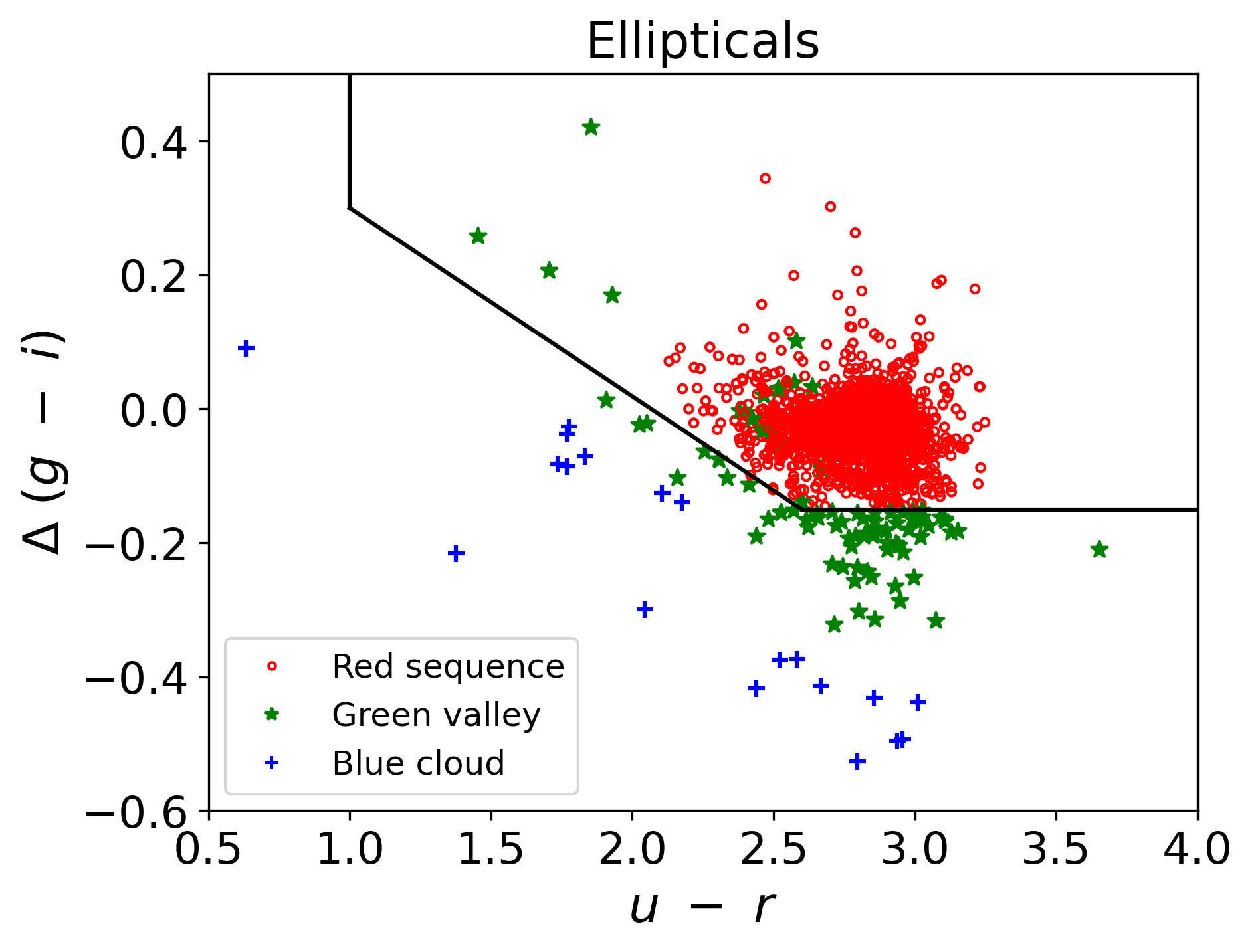}
    \caption{Color-magnitude diagram for the sample of spiral (top) and elliptical (bottom) galaxies. \citep[The top figure is based on Fig.~1 from ][]{Ravi24}. The open red circles are red-sequence galaxies, green stars are green-valley galaxies, and blue crosses are blue-cloud galaxies. The black line separates early- and late-type galaxies.}
    \label{fig:CMD}
\end{figure}

To investigate how the mass components of spiral and elliptical galaxies compare depending on the evolutionary stage, we separate them into three populations based on the color-magnitude diagram (CMD): the blue cloud, the green valley, and the red sequence. Blue-cloud galaxies are typically fainter, bluer, and forming stars, while red-sequence galaxies are brighter, redder, and have less star formation. The green valley galaxies are the transition between these two populations \citep{Martin07}. Hence, the CMD classification reflects the evolutionary stage of each galaxy.
 
We use the same CMD classification based on the $u-r$ color, the $\Delta(g-i)$ color gradient, and the inverse concentration index, $c_{\rm inv}$ as defined in \cite{Douglass22} and used in \cite{Ravi24}. Fig.~\ref{fig:CMD} shows the color-magnitude diagram for the spiral (top) and elliptical (bottom) galaxy samples. The black line defines the boundary between early-type galaxies, which fall above and to the right of the line, and late-type galaxies, which fall below and to the left of the line \citep{Park05}. Blue-cloud galaxies are defined as normal late-type galaxies and red-sequence galaxies are defined as normal early-type galaxies~\citep{Choi10}. Blue early-type galaxies, galaxies that lie above the boundary but with $u-r<2$ or a high $c_{\rm inv}$, and galaxies below the boundary but with $\theta < 20^\circ$ are classified as green-valley galaxies, where
\begin{equation}
    \tan \theta = \frac{-\Delta(g-i) + 0.3}{(u-r)-1} .
\end{equation}
The spiral galaxies populate all three CMD classifications, with the majority being in the blue cloud and green valley, while almost all elliptical galaxies are classified as red sequence. This agrees with predictions of hierarchical galaxy formation theories that spiral galaxies merge to form elliptical galaxies, which are at the end of the galaxy evolution track \citep{deLucia06}. In the following, we therefore do not separate the elliptical galaxies based on their CMD classification. 


\section{Estimating the mass of a galaxy}\label{sec:method}

In this section, we discuss the methodology of determining the baryonic and the total (dynamic, or gravitational) mass for both spiral and elliptical galaxies. We also introduce the effective rotational velocity of elliptical galaxies, which we later use to construct an extended BTFR for spiral and elliptical galaxies.

\subsection{The Baryonic Mass}\label{sec:bar_mass_components}

In the following, we describe the evaluation of the baryonic mass of galaxies, which we assume is composed of stars and gas. There are three main components to a galaxy's gas mass: neutral atomic hydrogen, \HI, molecular hydrogen, \Htwo, and helium, He, so we estimate the gas mass as the sum of these three components. To estimate the stellar mass of both spiral and elliptical galaxies, we use the stellar mass density maps from the Pipe3D MaNGA value added catalog \citep{Sanchez16, Sanchez18}.

\subsubsection{The Stellar Mass of Elliptical Galaxies}\label{sec:elliptical_stellar_mass}

\begin{figure}
    \centering
    \includegraphics[width=0.49\textwidth]{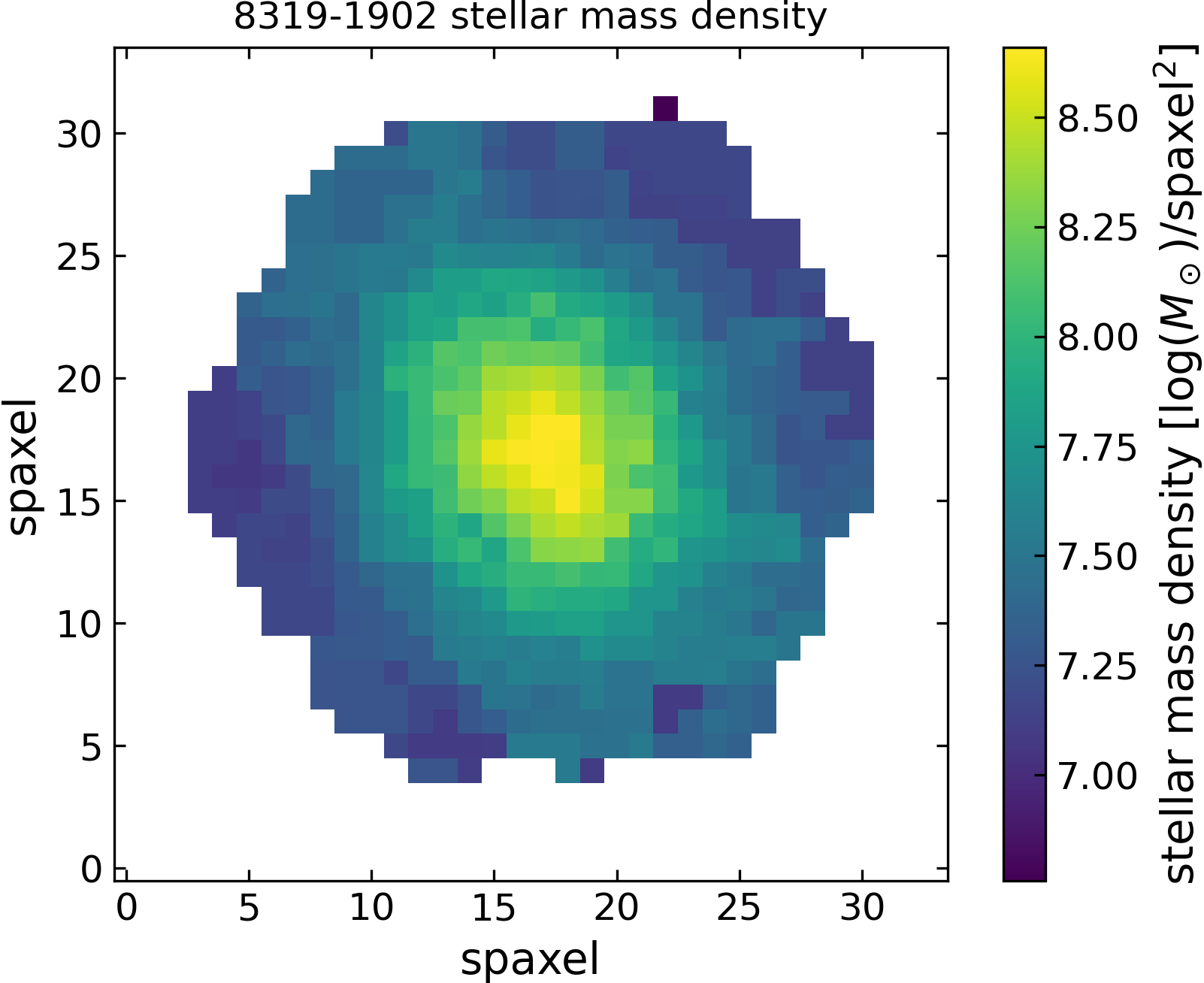}
    \includegraphics[width=0.49\textwidth]{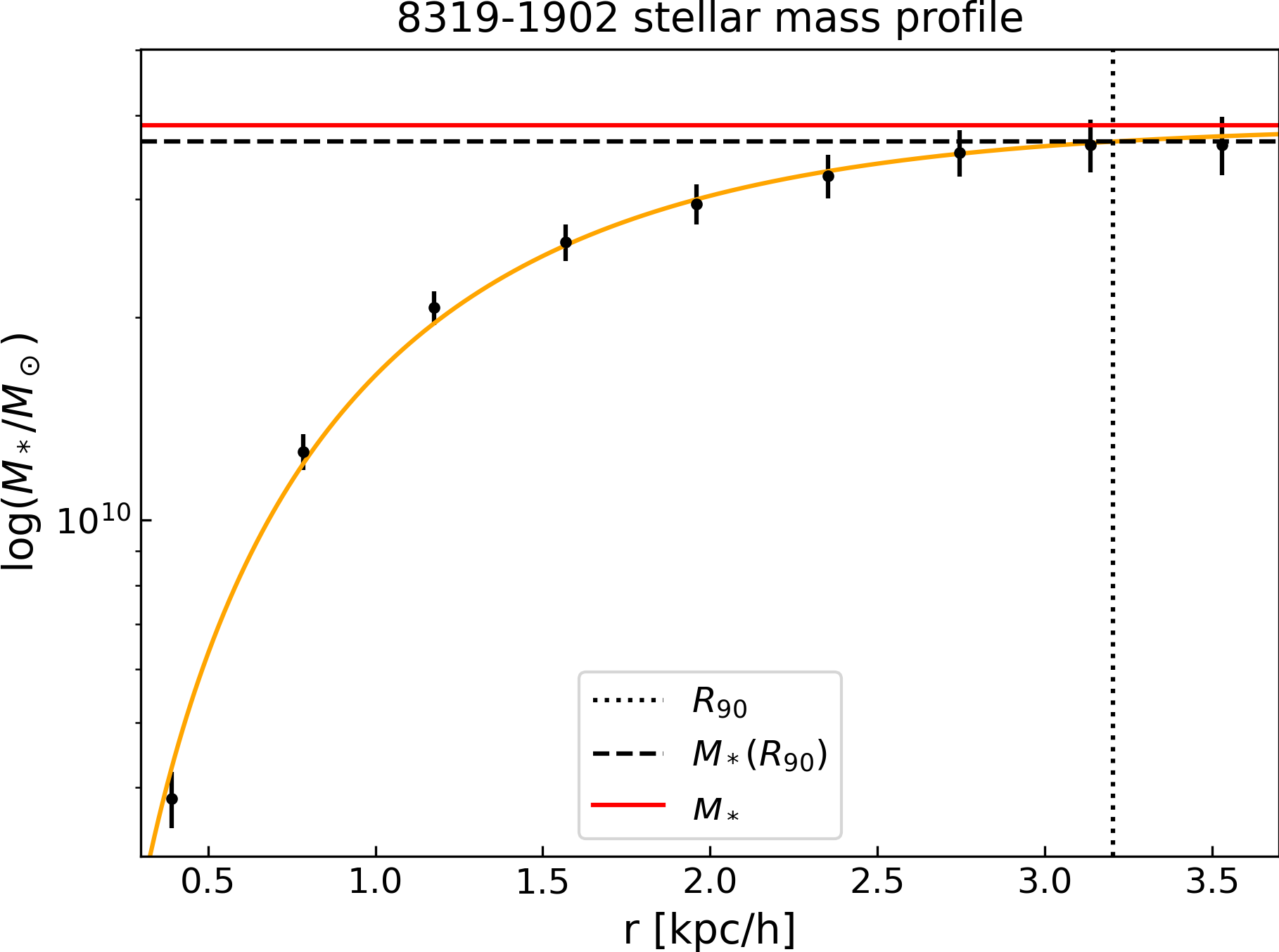}
    \caption{Example of stellar mass density map from Pipe3D (top) and best-fit exponential sphere and disk model to this map (bottom). In the bottom figure, the vertical dotted black line corresponds to $R_{90}$ for this galaxy, the dashed black horizontal line shows the stellar mass within $R_{90}$, and the solid horizontal red line shows the total stellar mass of the galaxy.}
    \label{fig:sMass_model}
\end{figure}

Using the stellar mass density map of each elliptical galaxy and the photometric inclination angle and position angle from the NSA, we draw ellipses at different radii from the brightest spaxel in the $g$-band weighted mean flux map, with the semi-major axis of each ellipse differing by 2 spaxels. By summing the stellar mass density of all spaxels within each ellipse, we obtain the stellar mass as a discretized function of radius, $M_*(r)$.

\begin{figure}
    \centering
    \includegraphics[width=0.49\textwidth]{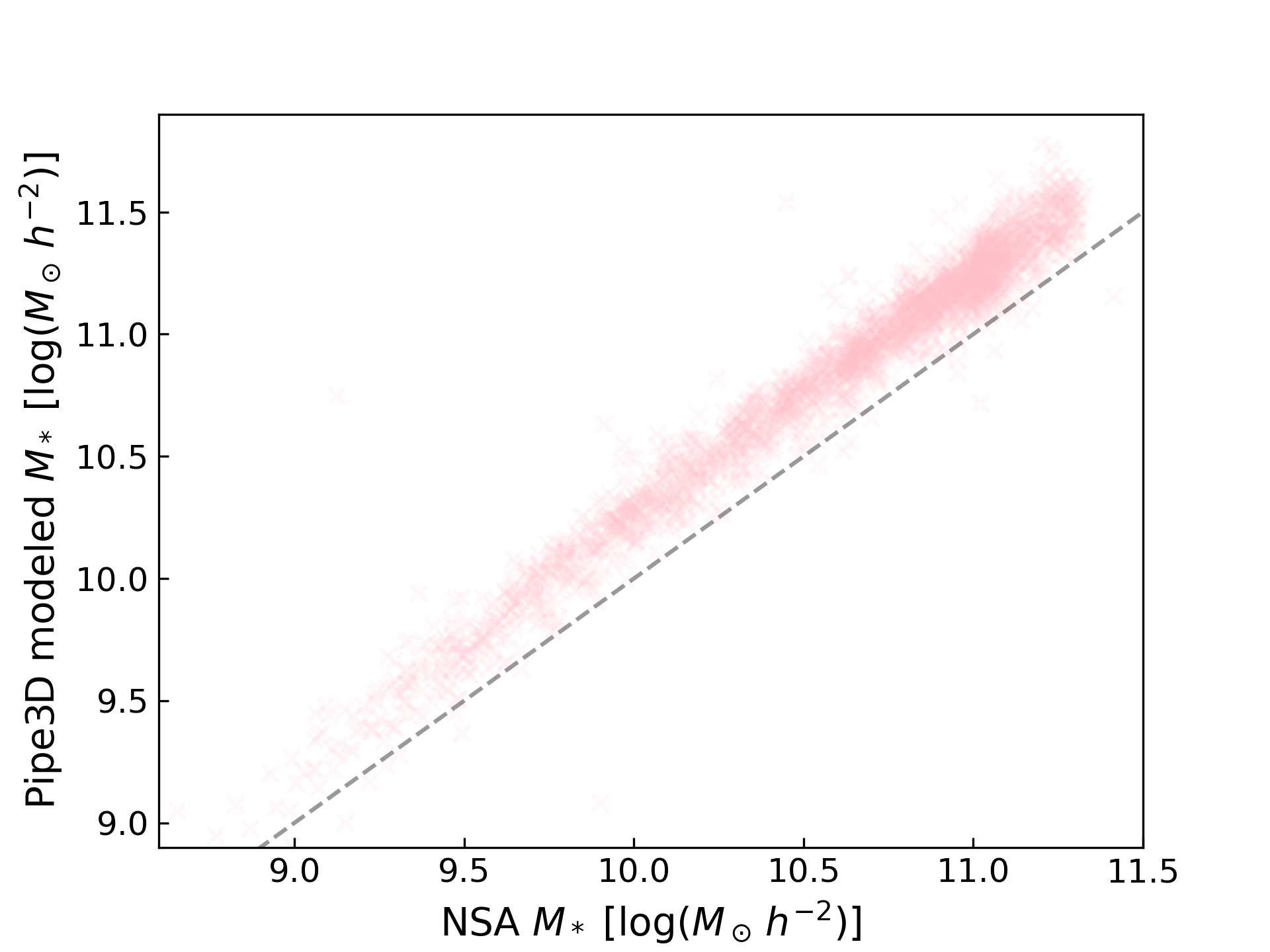}
    \caption{Comparison of stellar mass from the NSA and the total stellar mass, $M_*$ determined from the fit to Eq.~\ref{eq:smass}. The grey dashed line shows where the two masses are equal.}
    \label{fig:nsa_pipe3d_compare}
\end{figure}

These data are used to model the stellar mass, $M_*(r)$, within some radius $r$ as the sum of an exponential sphere (or bulge), $M_b(r)$, and an exponential disk, $M_d(r)$ \citep{Meert13, Sofue17}:
\begin{equation}\label{eq:smass_r90}
    M_*(r) = M_b(r) + M_d(r), 
\end{equation}
with the functional form of the components given by
\begin{equation}\label{eq:M_b}
    M_b(r) = M_0\,F(r/R_b),
\end{equation}
where $M_0 = 8 \pi R_b^3 \rho_b$ and $F(x) = 1 - e^{-x}(1+x+0.5x^2)$, and
\begin{equation}\label{eq:M_d}
    M_d(r) = 2\pi \Sigma_d R_d[R_d-e^{-r/R_d}(r+R_d)].
\end{equation}
The free parameters in the fit are $R_b$, the scale radius of exponential sphere, $\rho_b$, the central density of the exponential sphere, $R_d$, the scale radius of the disk, and $\Sigma_d$, the central surface density of the disk. Fig.~\ref{fig:sMass_model} shows an example of a stellar mass density map from Pipe3D and the corresponding model.

Using the parameters for this best-fit model, we compute $M_*(R_{90})$, the stellar mass within $R_{90}$, and the total stellar mass, 
\begin{equation}\label{eq:smass}
    M_* = M_0 + 2 \pi \Sigma_d R_d^2.
\end{equation}
Fig.~\ref{fig:nsa_pipe3d_compare} shows a comparison of the stellar masses for the elliptical galaxies obtained from the NSA and the stellar masses estimated using the described model. The stellar mass based on our fit is systematically higher by 0.25 dex. The derived stellar masses in the NSA are based on the Chabrier 2003 IMF \citep{Chabrier03}. Pipe3D estimates the stellar masses using the Salpeter 1955 IMFs \citep{Salpeter55}. The mean difference of 0.25 dex is close to $\log(0.55)$, which is the difference in stellar mass estimates based on Chabrier 2003 and Salpeter 1955 IMF derived by \cite{Longhetti09}. Hence, the observed difference is expected and explained by the difference in the IMF models.

\subsubsection{The Stellar Mass of Spiral Galaxies}
 
Similar to elliptical galaxies, the stellar mass of spiral galaxies is modeled as a sum of the bulge and disk components (Eqs.~\ref{eq:smass_r90}--\ref{eq:M_d}). However, for spiral galaxies, we follow \cite{Ravi24} and model the rotation curve of the stellar component to find the total stellar mass.  
The rotational velocity corresponding to the bulge model (Eq.~\ref{eq:M_b}) is \citep{Sofue17}
\begin{equation}
    V_b(r)^2 = \frac{GM_0}{R_b}F(r/R_b),
\end{equation}
where $G = 6.67408 \times 10^{-11}$ m$^3$ kg$^{-1}$ s$^{-2}$ is the gravitational constant. The rotational velocity due to the disk model (Eq.~\ref{eq:M_d}) is 
\begin{equation}
    V_d(r)^2 = 4\pi G \Sigma_d R_d y^2[I_0(y)K_0(y) - I_1(y)K_1(y)],
\end{equation}
where $y=r/2R_d$, and $I_i$ and $K_i$ are the modified Bessel functions. The total rotational velocity, $V_*$, is the sum of these components in quadrature
\begin{equation}
    V_*(r)^2 = V_b(r)^2 + V_d(r)^2.
\end{equation}
 
The rotational velocities due to the stellar mass densities are fit to $V_*(r)$ with $\rho_b$, $R_b$, $\Sigma_d$, and $R_d$ as free parameters, using the stellar mass density map and the updated values for the position angle, inclination, and galaxy center obtained from the fits described in Sec.~\ref{sec:Spiral_fit}. The stellar mass within $R_{90}$ and the total stellar mass for the spiral galaxies are evaluated according to Eqs.~\ref{eq:smass_r90} and \ref{eq:smass}, respectively. For galaxies where the fit results in $R_b > 10$~kpc, we set the stellar mass equal to only the mass of the disk component, as this bulge size is unphysically large.

\subsubsection{The Gas Mass}\label{sec:gas_mass}

\begin{figure}
    \centering
    \includegraphics[width=0.49\textwidth]{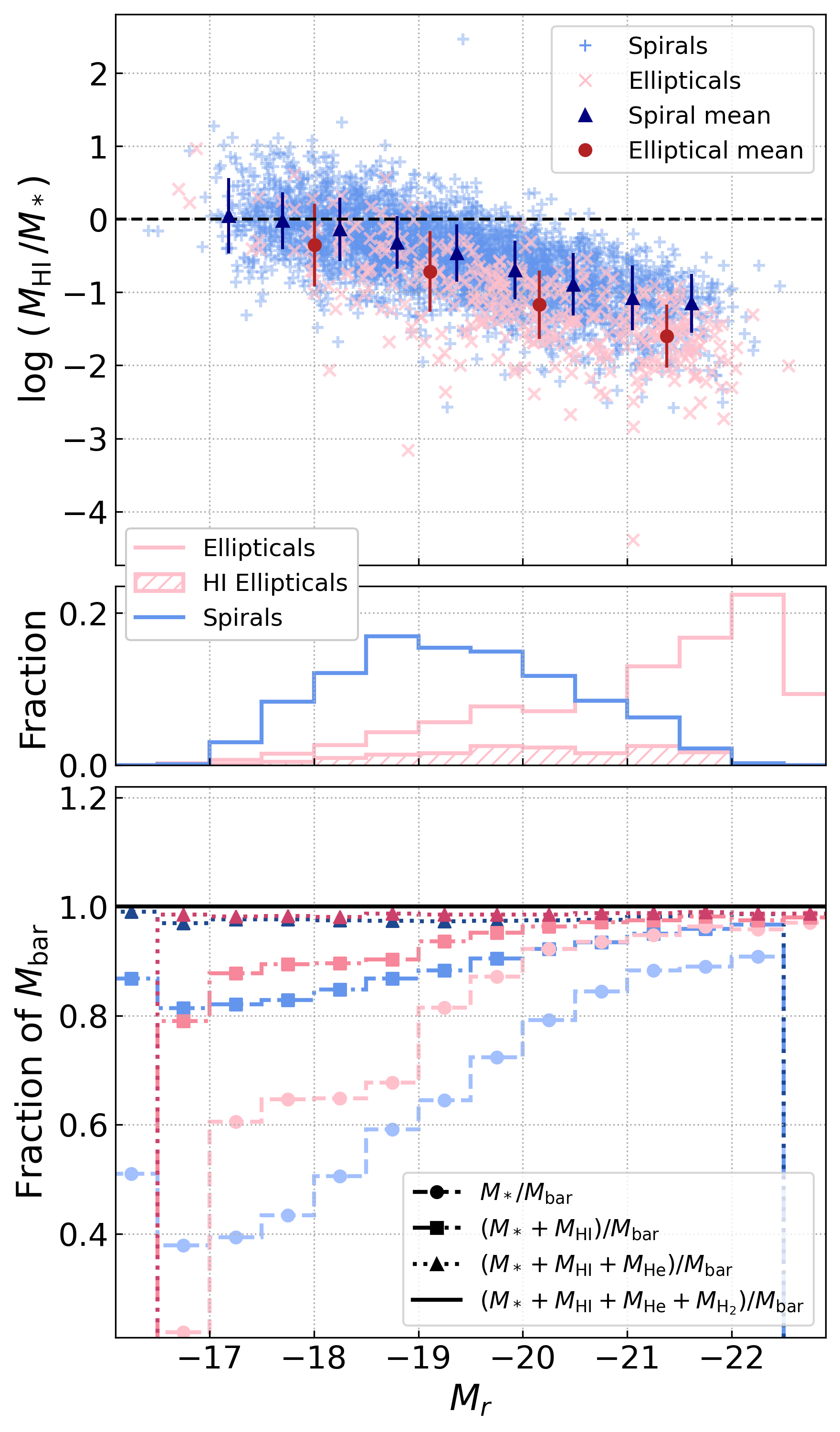}
    \caption{\emph{Top:} The ratio of \HI mass to stellar mass as a function of $M_r$ for the MaNGA spirals (blue crosses) and ellipticals (pink x's). The navy triangles (spirals) and red circles (ellipticals) are the means from fits to a Gaussian distribution in bins of $M_r$. The black dashed line shows where the stellar mass equals the \HI mass. \\
    \emph{Middle:} The $M_r$ distribution of all galaxies (open), and those with \HI detections (hatched histogram).\\  
    \emph{Bottom:} The fractions of the baryonic mass components as a function of $M_r$. 
    }
    \label{fig:ellipticals_HI}
\end{figure}

Gas mass is comprised of three dominant components: \HI, \Htwo, and He. Information about the \HI mass is obtained from the \HI--MaNGA DR4 survey \citep{Masters19, Stark21}, and is shown in Fig.~\ref{fig:ellipticals_HI} as a function of $M_r$ for the spiral and elliptical galaxies. Out of our sample of 2460 elliptical galaxies, 1034 were observed in the \HI--MaNGA survey, 652 of which were non-detections; hence, \HI mass is available for only 15.5\% of the elliptical galaxy sample. The top panel of Fig.~\ref{fig:ellipticals_HI} shows the ratio of \HI mass to stellar mass as a function of luminosity. The \HI mass of ellipticals follows the same trend as spiral galaxies, where the \HI fraction decreases with increasing luminosity, but in ellipticals it is lower than in spirals at all luminosities. We note that most elliptical galaxies with \HI detections have \HI masses $\lesssim 0.1M_*$. Since the majority of the elliptical galaxies have higher luminosities, as shown in the middle panel of Fig.~\ref{fig:ellipticals_HI}, we can safely neglect the contribution of gas to the baryonic mass of elliptical galaxies and only consider their stellar mass. Thus, the full sample of 2460 elliptical galaxies is retained for further analysis. 

The \Htwo mass is estimated using the parameterization as a function of $M_r$ and CMD classification from \cite{Ravi24} based on the MASCOT survey, the xCOLD GASS survey, and SDSS~DR7. We estimate the He mass as a 25\% mass fraction of the total H gas content \citep{Cooke18}.

Using these parameterizations, we compare the contributions of each mass component to the total baryonic mass of elliptical and spiral galaxies in the bottom panel of Fig.~\ref{fig:ellipticals_HI}. For the brighter elliptical galaxies, which constitute the majority of the sample, the stellar mass is $\gtrsim 0.9M_{\rm bar}$. In contrast, the gas, particularly \HI and He, is a significant part of the baryonic mass for spiral galaxies, especially those with low luminosity. Therefore, from the 5626 spiral galaxies, we use the 3283 that include \HI observations for further analysis.

\subsubsection{Calculating the Baryonic Mass}

The baryonic mass for a spiral galaxy is evaluated as the sum of its stellar, \HI, \Htwo, and He masses:
\begin{equation}
    M_{\rm bar} = M_* + \text{\MHI} + M_{\rm H_2} + M_{\rm He}.
\end{equation}
As described in Sec.~\ref{sec:gas_mass}, the gas content of elliptical galaxies is negligible, so their baryonic mass is just equal to the stellar mass:
\begin{equation}
    M_{\rm bar} = M_*.
\end{equation}

\subsection{The Total Mass}\label{sec:total_mass}

We estimate the total mass of a galaxy from its kinematics in order to probe its gravitational potential. For spiral galaxies, this involves modeling the rotation curve of the galaxy, while for ellipticals, we use the stellar velocity dispersion to estimate the total mass.

\subsubsection{The Total Mass of Spiral Galaxies} \label{sec:Spiral_fit}

To estimate the total mass of spiral galaxies, we follow the procedure in \cite{Ravi24}, where each galaxy's rotational velocity at a radius $r$ is fit to the parameterization \citep{BarreraBallesteros18}
\begin{equation}\label{eq:BB}
    V(r) = \frac{V_{\rm max}r}{(R_{\rm turn}^\alpha + r^\alpha)^{1/\alpha}},
\end{equation}
where $V_{\rm max}$, the velocity at which the rotation curve reaches the plateau, $R_{\rm turn}$, the radius at which the rotation curve flattens, and $\alpha$, the sharpness of the curve, are free parameters in the fit. The systemic velocity, kinematic center, position angle, and inclination are also free parameters, but unlike \cite{Ravi24}, the inclination angle is restricted to within 15$^\circ$ of the inclination calculated based on the photometric axis ratio from the NSA. We select the best fit model as the one with the minimum $\chi^2 = \Sigma((\text{data} - \text{model})/\text{uncertainty})^2$.

The total mass within a radius $r$ is calculated as
\begin{equation}\label{eq:spiral_mtot}
    M(r) = \frac{V(r)^2r}{G}.
\end{equation}
In the following, we use the total mass of each spiral galaxy within $R_{90}$, \Mtot $= M(R_{90})$.

\subsubsection{Elliptical Galaxy Total Mass}\label{sec:elliptical_mass}

\begin{figure*}
    \centering
    \includegraphics[height=0.2\textheight]{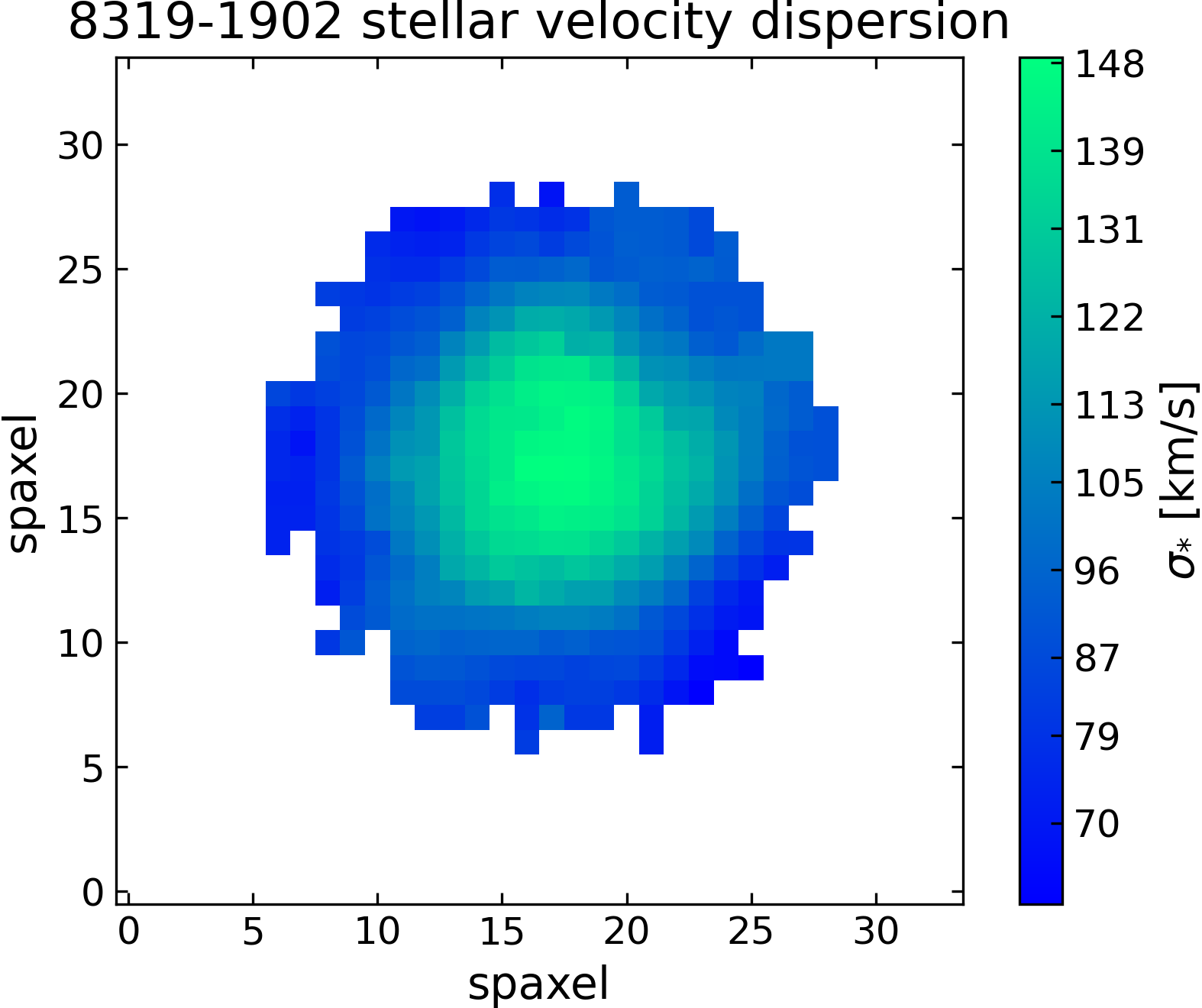}
    \includegraphics[height=0.2\textheight]{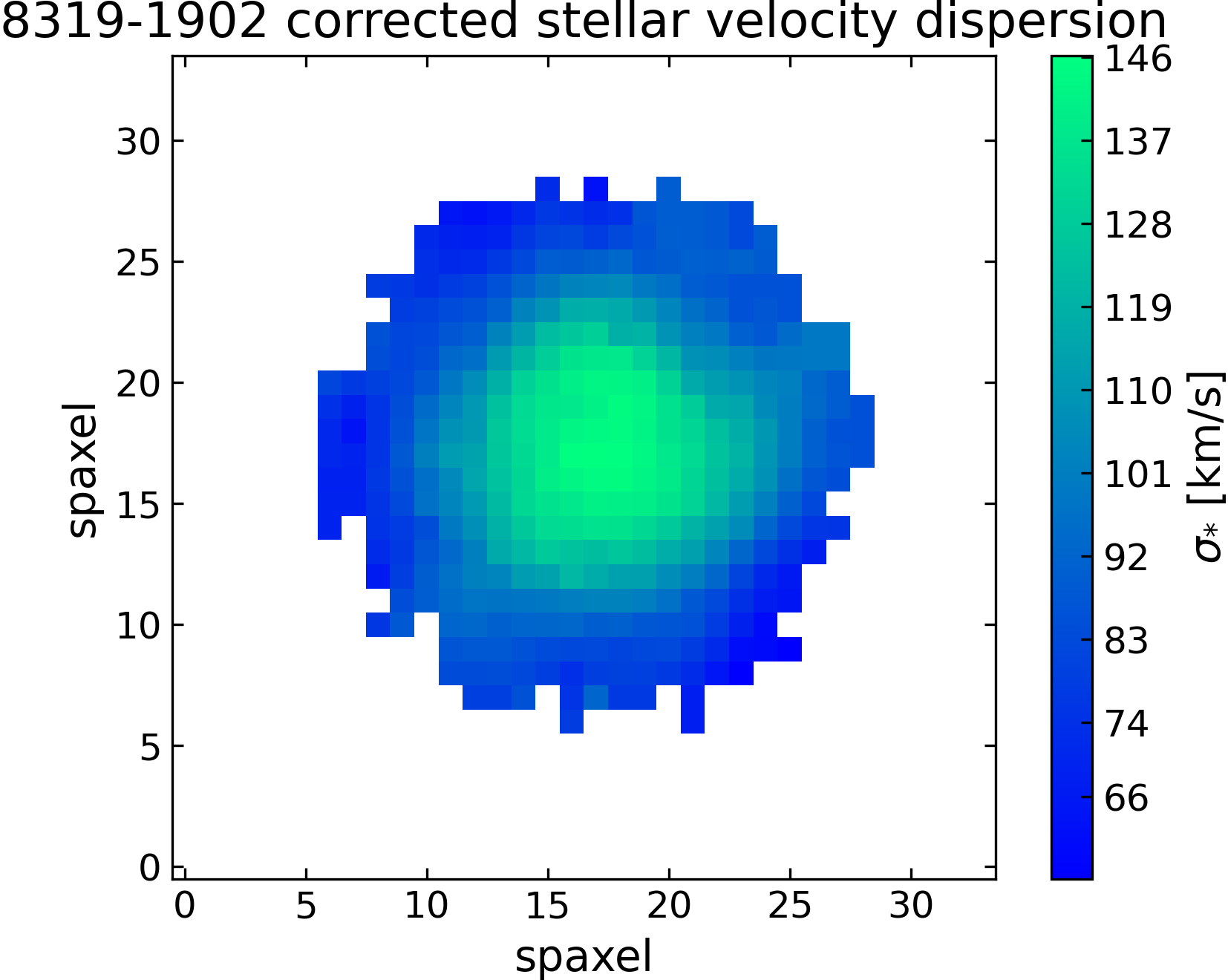}
    \includegraphics[height=0.2\textheight]{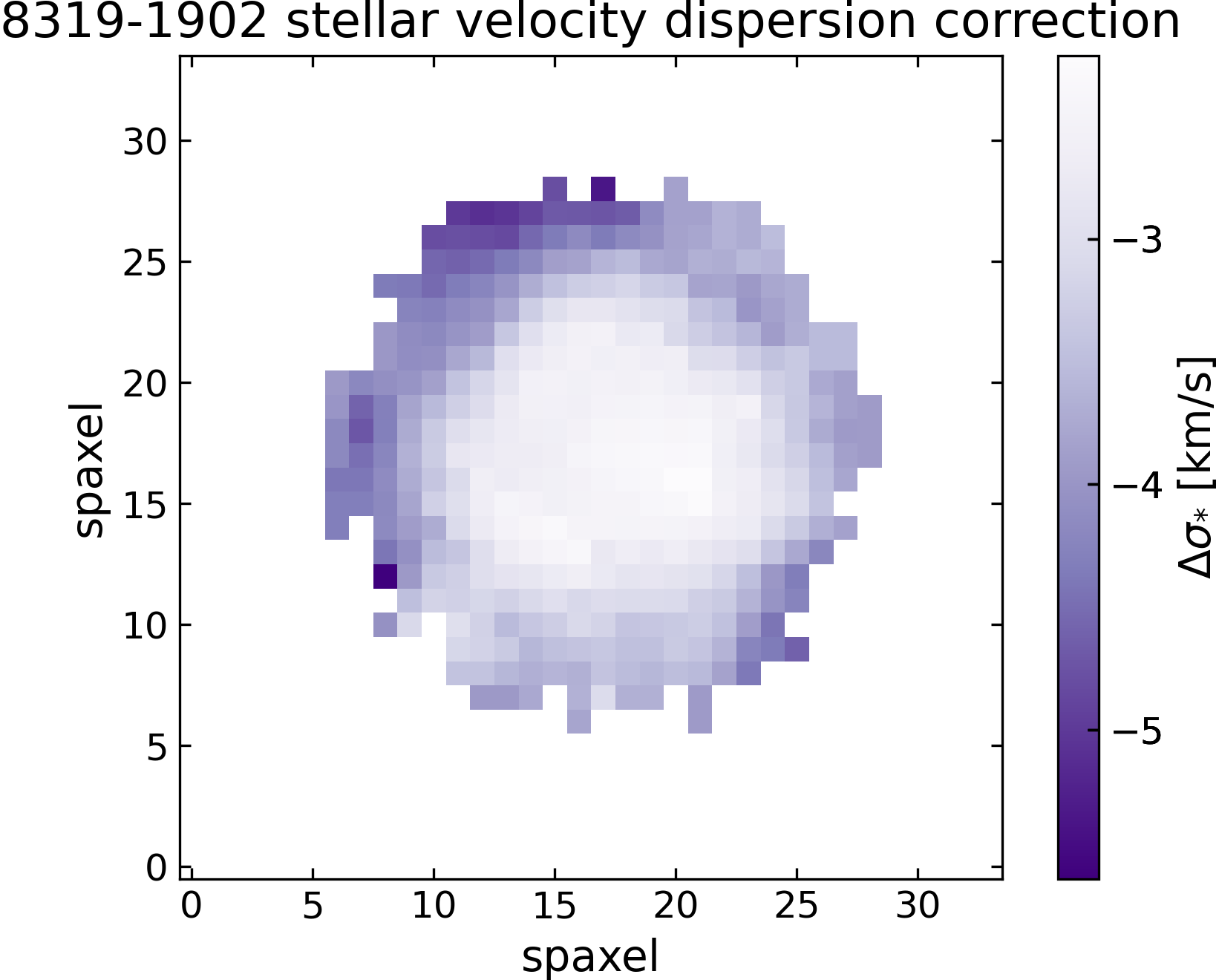}
    \caption{Example of the stellar velocity dispersion map from the MaNGA DAP (left), the same map corrected for instrumental effects (center), and the difference between the two (right).}
    \label{fig:vel_disp}
\end{figure*}

To calculate the total mass of each elliptical galaxy, we use the stellar mass density map from Pipe3D and stellar velocity dispersion map from the SDSS MaNGA DAP. The DAP also includes data quality bits and uncertainties on the stellar velocity dispersion map, and the correction factor for the instrumental dispersion, $\delta \sigma_{\rm inst}$. From both maps, we remove spaxels with either a nonzero data quality bit or a signal-to-noise ratio less than 10.

The stellar velocity dispersion, corrected for the instrument resolution, is evaluated from the observed dispersion in each spaxel, $\sigma_{\rm obs}$:
\begin{equation}
    \sigma^2 = \sigma_{\text{obs}}^2 - \delta \sigma_{\text{inst}}^2 .
\end{equation}
In Fig.~\ref{fig:vel_disp}, we show an example stellar velocity dispersion map, the same map with the correction applied, and the difference between the two maps. We then calculate the stellar-mass-weighted mean velocity dispersion
\begin{equation}
    \sigma_* = \frac{\sum_i M_{*,i}\sigma_{i}}{\sum_i M_{*,i}} ,
\end{equation}
where $M_{*,i}$ is the stellar mass in spaxel $i$.

An elliptical galaxy has little to no rotation, with its stars moving randomly. Thus, assuming thermal equilibrium, we apply the virial theorem to calculate the total mass of a galaxy \citep{Ryden20}:
\begin{equation}\label{eq:vir_thm}
    M_{\rm tot} = 7.5 \frac{\sigma_*^2\,R_{50}}{G}, 
\end{equation}
where we use the half-light radius, $R_{50}$, as a proxy for the half-mass radius.

\subsubsection{Effective Rotational Velocity}\label{sec:effectiveV}

In order to extend the BTFR to elliptical galaxies, we construct an ``effective rotational velocity,'' the rotational velocity an elliptical galaxy would have at $R_{90}$ if it were dominated by rotational motion instead of random motion:
\begin{equation}\label{eq:veff}
    V_{\rm eff} = \sqrt{\frac{G M_{\rm tot}}{R_{90}}},
\end{equation}
where $M_{\rm tot}$ is given by Eq.~\ref{eq:vir_thm}.

\section{Verification of the Methodology with TNG100 Galaxies}\label{sec:verification}

\subsection{Matching Observed and Simulated Galaxies} \label{sec:matching}

\begin{figure}
    \centering
    \includegraphics[width=0.49\textwidth]{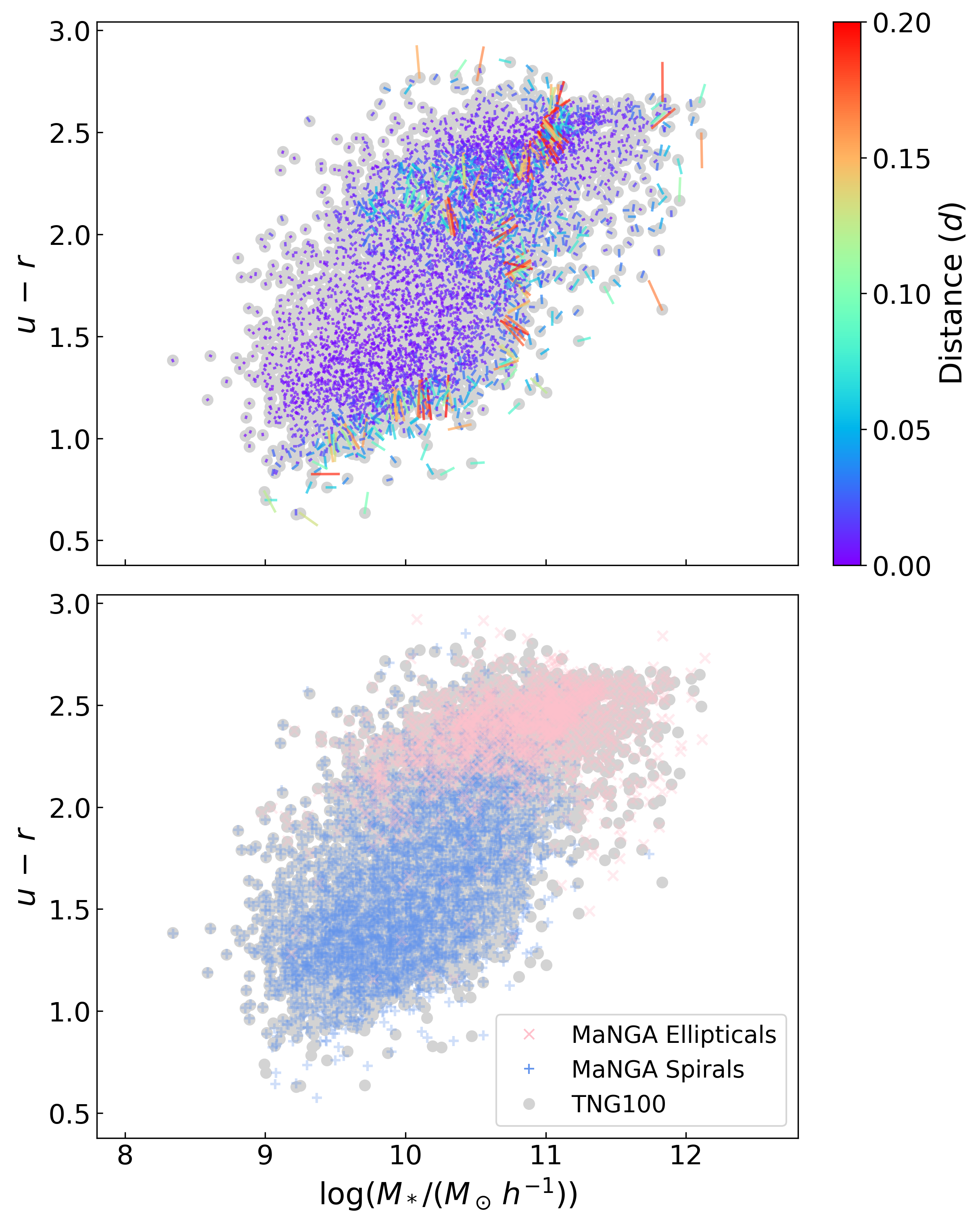}
    \caption{Color--stellar mass diagram of the MaNGA and matched TNG100 galaxies. The grey points are the TNG100 galaxies. \emph{Top:} The color-coded line segments show the distance between each MaNGA--TNG100 matched galaxy pair. \emph{Bottom:} MaNGA spiral (blue crosses) and elliptical (pink x's) galaxies over TNG100 galaxies.}
    \label{fig:manga_tng_matching}
\end{figure}

In order to verify our methodology and compare the BTFR for MaNGA galaxies to predictions from $\Lambda$CDM, we select a sample of galaxies in the IllustrisTNG 100-1 (TNG100) simulation that match our sample of MaNGA galaxies using $u-r$ color and stellar mass. We use the synthetic SDSS colors from the simulation galaxies \citep{Nelson18} and compare those to the NSA $u-r$ color for the MaNGA sample. 

As discussed in Sec~\ref{sec:elliptical_stellar_mass} the difference in the IMF used in TNG100 and Pipe3D leads to a stellar mass difference of $\log(0.55)$. We thus convert TNG100 stellar masses to the Salpeter IMF basis using this factor before matching. 
We match the galaxies by minimizing the distance $d$ in the $(u-r)$--$M_*$ space:
\begin{equation}\label{eq:match}
  d^2 = (M_{*, \rm sim} - M_{*, \rm obs})^2 + ((u-r)_{\rm sim} - (u-r)_{\rm obs})^2.
\end{equation}
Here, quantities with a subscript ``sim'' denote values from the TNG100 simulation, and quantities with a subscript ``obs'' denote values from the MaNGA observations. We require that $d<0.2$ for a valid match.

For each MaNGA galaxy, we take the closest simulation galaxy in the $(u-r)$--$M_*$ space. If multiple MaNGA galaxies match to the same simulation galaxy, we match the simulated galaxy to the MaNGA galaxy with the smallest $d$. We then remove these matched MaNGA and simulation galaxies from the sample and repeat the matching process on the remaining MaNGA galaxies. This is done until there are no more matches within a distance of 0.2. Fig.~\ref{fig:manga_tng_matching} shows the matched MaNGA and TNG100 galaxies in the $(u-r)$--$M_*$ plane.

This matching results in a final sample of 3149 spiral galaxies and 1423 elliptical galaxies for the BTFR analysis. Not every MaNGA galaxy has a corresponding TNG100 galaxy, since there are not enough galaxies in TNG100 within this $(u-r)$--$M_*$ range due to its limited volume. Table~\ref{tab:sample_cuts} shows the number of spiral and elliptical galaxies after each step of the analysis.

\begin{deluxetable}{lCC}
    \tablecaption{BTFR Sample Counts}
    \label{tab:sample_cuts}
    \tablehead{
        \colhead{Cut} & \colhead{Spirals} & \colhead{Ellipticals}}
    \startdata
        Smoothness and T-type & 5626 & 2460 \\
        HI observations & 3283 & 2460 \\
        TNG100 match & 3149 & 1423\\
    \enddata
\end{deluxetable}

\subsection{Comparison with TNG100 Galaxies}\label{sec:total_mass_comparison_sim}

\begin{figure*}
    \centering
    \includegraphics[width=0.95\textwidth]{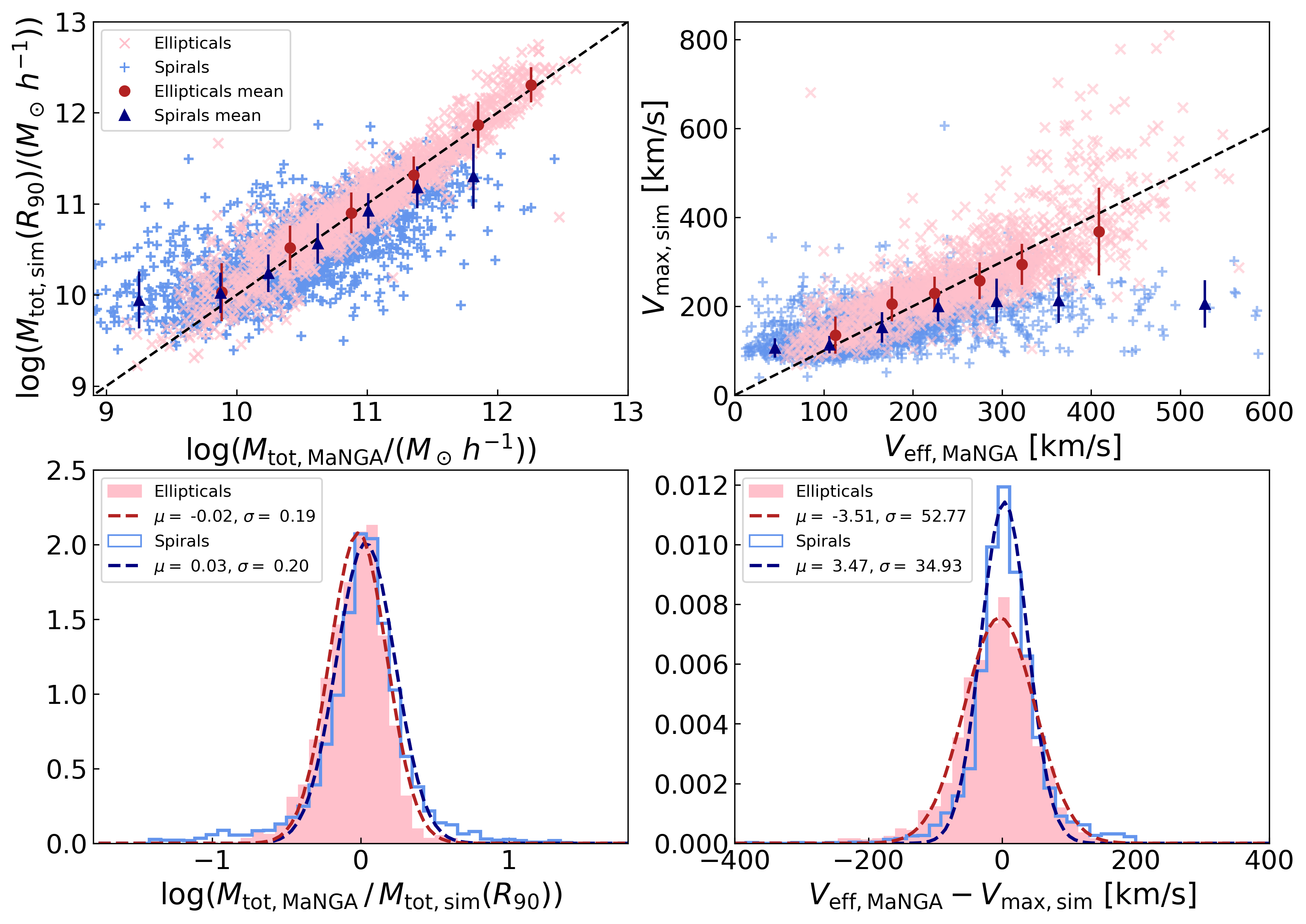}
    \caption{\emph{Top:} Comparison of total mass (left), and the rotational velocity (right) of MaNGA spiral (blue crosses) and elliptical (pink x's) galaxies and their matched TNG100 galaxies. The black dashed lines show where masses or velocities are equal. The red circles and navy triangles show the means from the Gaussian fit to the distribution of each mass or velocity bin for ellipticals and spirals respectively. 
    \emph{Bottom:} The PDF of the difference between MaNGA galaxies and their matched TNG100 galaxies. The results of a fit to a Gaussian distribution are shown in dashed lines and given in the inserts.}
    \label{fig:tng_compare}
\end{figure*}

\begin{figure}
    \centering
    \includegraphics[width=0.49\textwidth]{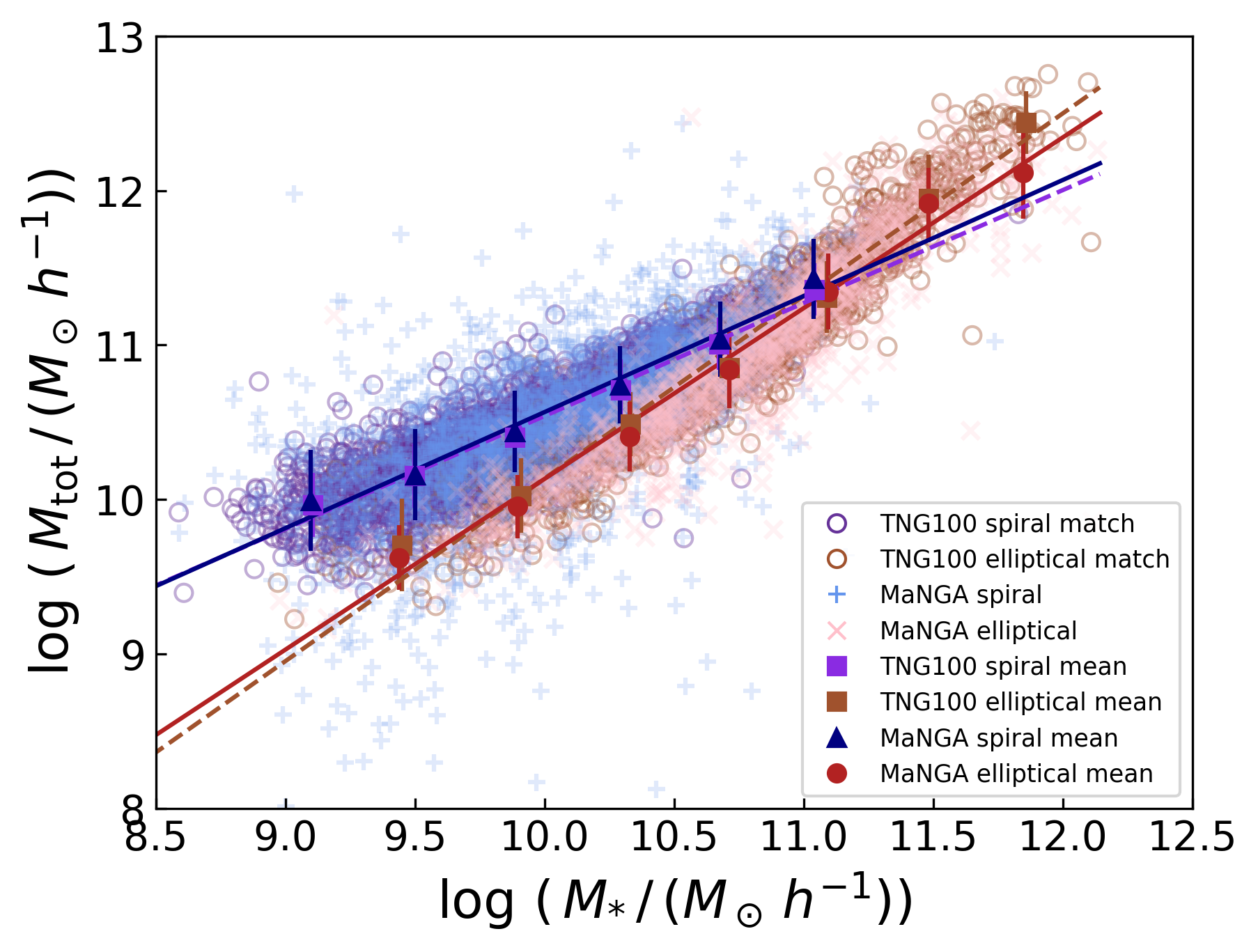}
    \caption{Stellar vs.~total mass for the MaNGA spiral (blue crosses) and elliptical (pink x's) galaxies and matched TNG100 spiral (purple circles) and elliptical (brown circles) galaxies. The navy triangles (MaNGA spirals), red circles (MaNGA ellipticals), purple squares (TNG100 spirals) and brown squares (TNG100 ellipticals) show the mean from fitting each stellar mass bin to a Gaussian distribution. Linear fits are shown in the purple (spiral) and brown (elliptical) dashed lines for TNG100 galaxies and navy (spiral) and red (elliptical) solid lines for MaNGA galaxies. See Table~\ref{tab:mstar_mtot_tng_fits} for the fit results. 
    }
    \label{fig:manga_tng_mtot_mstar}
\end{figure}

To confirm that our estimates of the total mass, rotational velocity, and effective velocity of the MaNGA galaxies are accurate representations of the true total mass, rotational velocity, and effective velocity, we compare the MaNGA galaxies to their matched TNG100 galaxies. For the MaNGA sample, the total mass within $R_{90}$ is defined using Eq.~\ref{eq:spiral_mtot} for the spiral and Eq.~\ref{eq:vir_thm} for the elliptical galaxies. For each TNG100 galaxy in the matched sample, we take the maximum rotational velocity in its halo as $V(R_{90})$ and use the value of $R_{90}$ for the matched MaNGA galaxy to calculate $M(R_{90})$ using Eq.~\ref{eq:spiral_mtot}. 

In Fig.~\ref{fig:tng_compare} (left), we show the comparison of the total masses of the MaNGA spiral (blue) and elliptical (red) galaxies to that of the matched TNG100 galaxies. In the bottom plot we show the difference between the observations and the simulation, and fit these residuals to a Gaussian distribution. The mean and dispersion from the fit are shown in the inserts. We find good agreement for both spiral and elliptical galaxies. For elliptical galaxies, this observed agreement in the total mass shows that the application of the virial theorem with the half-light radius evaluates the total mass within $R_{90}$, \emph{not} the total mass of the galaxy. For spiral galaxies, the masses agree well for the majority of the population, with some deviations observed in the highest and lowest mass bins. Overall, the precision in the total mass evaluation is comparable between the spiral and elliptical galaxies. 

To verify that the definition of the effective velocity given in Eq.~\ref{eq:veff} is reasonable, we compare it to the maximum rotational velocity in the TNG100 galaxy halos for those matched to MaNGA ellipticals, as shown on the right in Fig.~\ref{fig:tng_compare}. We observe a very good agreement, which justifies using the effective velocity to extend the BTFR to elliptical galaxies. We also include the rotational velocity at $R_{90}$ for the observed spiral galaxies and maximum rotational velocity for the matched simulation galaxies. We find larger differences between the velocities for those galaxies with very high rotational velocities, which constitute only a small fraction of the sample. This may be due to the degeneracy between rotational velocity and galaxy inclination in the velocity map model for the MaNGA spirals. However, the majority of the sample shows a good agreement with the simulation galaxies, with comparable precision between the spiral and elliptical galaxies.

\begin{deluxetable}{lCC}
    \tablecaption{Parameters of the linear fits ($\log M_{\rm tot}(R_{90}) = a \log M_* + b$) from Fig.~\ref{fig:manga_tng_mtot_mstar}.
    \label{tab:mstar_mtot_tng_fits}}
    \tablehead{
        \colhead{Sample} & \colhead{$a$} & \colhead{$b$}}
    \startdata
        TNG100 spirals & 0.731 \pm 0.037 & 3.227 \pm 0.381\\
        MaNGA spirals & 0.751 \pm 0.047 & 3.051 \pm 0.484 \\
        TNG100 ellipticals & 1.182 \pm 0.063 & -1.688 \pm 0.683\\
        MaNGA ellipticals & 1.106 \pm 0.052 & -0.923 \pm 0.550
    \enddata
\end{deluxetable}

In Fig.~\ref{fig:manga_tng_mtot_mstar}, we show the $M_*$--\Mtot relation for the spiral and elliptical MaNGA and matched TNG100 galaxies. We quantify these relations with linear fits, the results of which are given in Table~\ref{tab:mstar_mtot_tng_fits}. We find that the slopes and $y$-intercepts agree well within uncertainties between data and simulation for both elliptical and spiral galaxies. This agreement justifies the methods that we use for estimating the total masses of the elliptical and spiral galaxies. We also note that the elliptical galaxies exhibit a tighter stellar-to-total mass relationship compared to the spiral galaxies, where the scatter increases for low mass galaxies. This behavior could be explained by a higher gas content, whose contribution to the total mass could vary, and a variation in the dark matter fraction in these galaxies.

\begin{figure}
    \centering
    \includegraphics[width=0.4\textwidth]{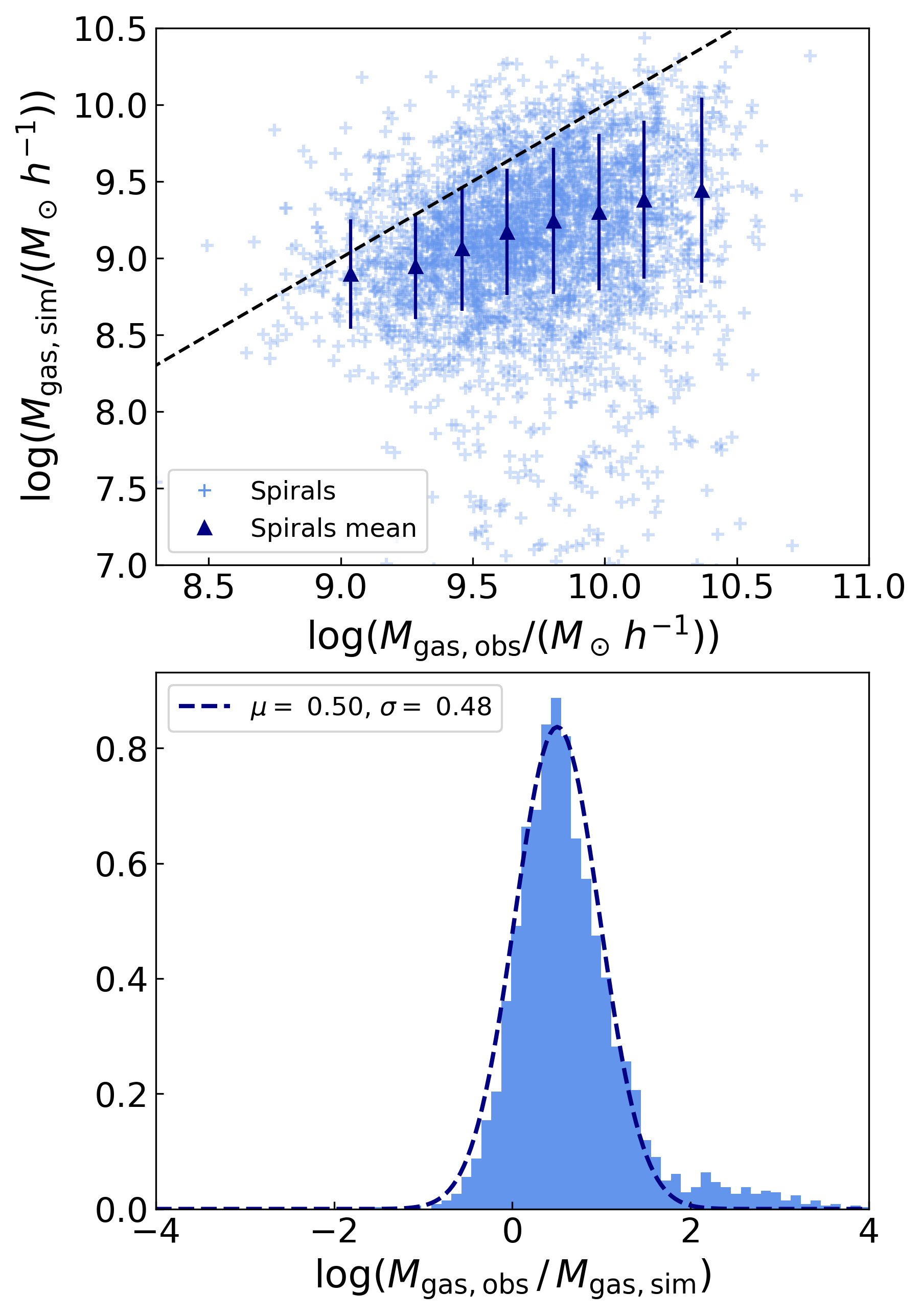}
    \caption{Comparison of the gas mass for MaNGA and matched TNG100 spiral galaxies. \emph{Top:} the blue crosses are individual galaxies, the navy triangles are the means in bins of $M_{\rm gas, obs}$, and the black dashed line shows where the gas mass from observations and simulation are equal. \emph{Bottom:} the PDF of the difference in the gas mass between data and simulation, with a Gaussian fit shown in the dashed line.} 
    \label{fig:gas_compare}
\end{figure}

In Fig.~\ref{fig:gas_compare}, we present a comparison of the gas mass observed in MaNGA to that in the matched TNG100 spiral galaxies. For the MaNGA spirals, we use the total observed \HI mass and the parameterized \Htwo and He mass, while for TNG100, we use the hydrogen and helium mass within twice the stellar half-mass radius of each galaxy, where the gas composition is provided by the simulation. The \HI disk in galaxies has been observed to extend beyond this radius \citep{Wang13}, and may extend past this radius in the simulation, as a result potentially underestimating the total \HI contribution to the baryonic mass in a given halo. The bottom panel of Fig.~\ref{fig:gas_compare} shows a Gaussian fit to the difference between the two gas estimates. We find that the gas masses obtained from the simulation are an underestimate of the observed gas mass.

\section{The Effect of an Evolving Stellar Population}\label{sec:baryonic_mass_comparison}

\begin{figure*}
    \centering
    \includegraphics[width=0.99\textwidth]{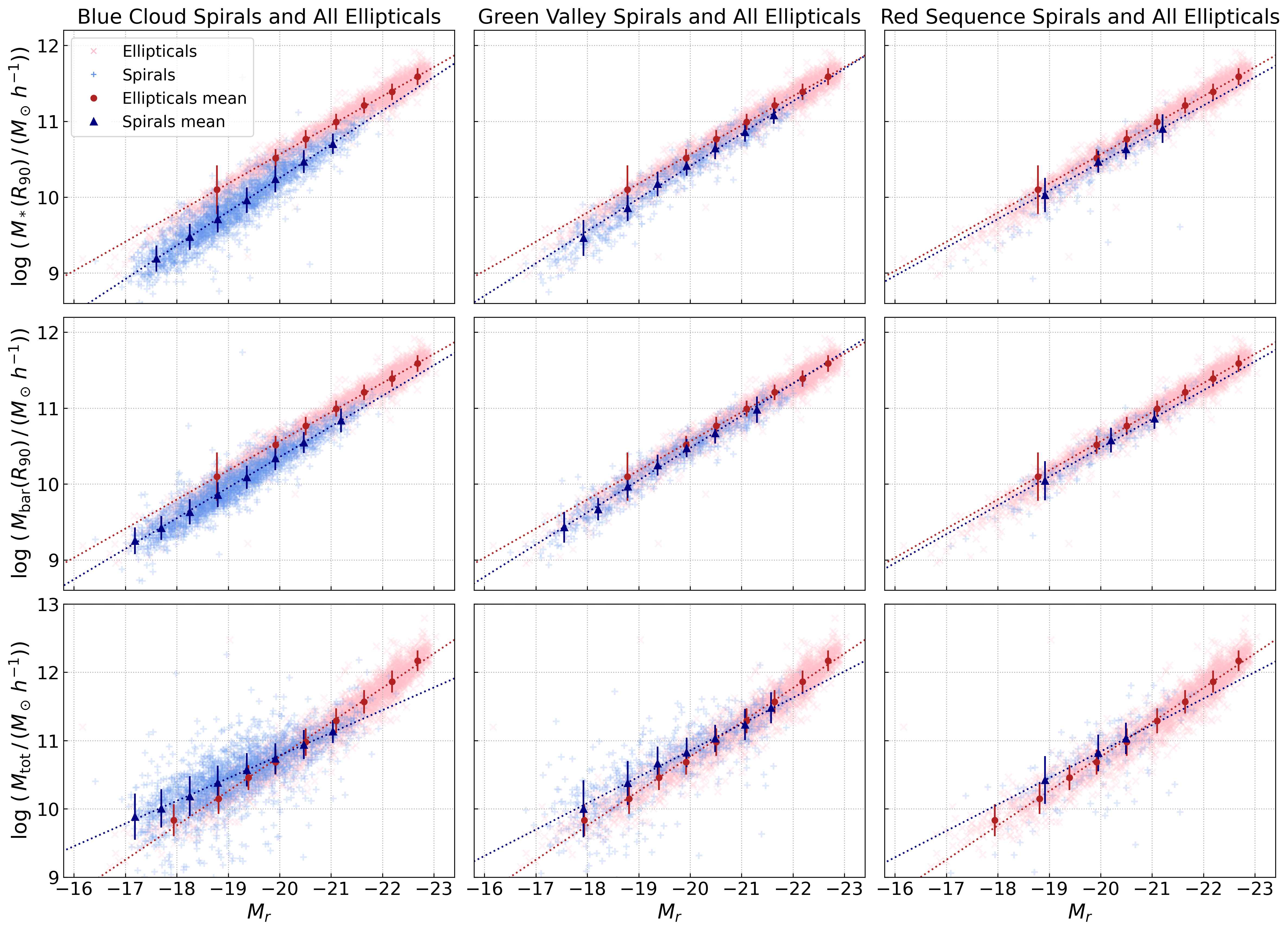}
    \caption{Stellar mass within $R_{90}$ (top), baryonic mass (middle), and total mass (bottom) as a function $M_r$ for all elliptical, blue-cloud (left), green-valley (center), and red-sequence (right) spiral galaxies. Blue crosses represent spiral galaxies, pink x's are elliptical galaxies, and navy triangles and red circles are means from fits to a Gaussian distribution for each $M_r$ bin for spirals and ellipticals respectively. See Table \ref{tab:fit_masses_Mr} for the fit results.}
    \label{fig:mtot_mstar_mr_all}
\end{figure*}

In Fig.~\ref{fig:mtot_mstar_mr_all}, we compare the stellar (top), baryonic (middle), and total (bottom) masses of elliptical and spiral galaxies separated by their CMD classification into the blue cloud (left), the green valley (center), and the red sequence (right) populations. As shown in Fig.~\ref{fig:CMD}, most of the elliptical galaxy sample lies in the red sequence category, and as a result we do not separate them by their CMD classification. The dependencies on $M_r$ for each population are fit to linear functions, with slopes and $y$-intercepts given in Table~\ref{tab:fit_masses_Mr}.

We find that the absolute values of the slopes of the stellar mass dependence on $M_r$ decrease along the evolutionary track from blue-cloud to green-valley to red-sequence spirals, with red sequence spirals and ellipticals having consistent slopes. In contrast, we do not see a slope change along the evolutionary track when all the baryonic mass components are considered. The slope of the baryonic mass--luminosity relation for the red-sequence and green-valley spirals is the same as the slope of the stellar mass--luminosity relation; the contribution of gas to the baryonic mass of these galaxies is smaller than in the blue cloud. The absolute values of the slopes of the \Mtot--$M_r$ relation increase along the evolutionary track. 

Overall, we find that there is a strong correlation in the stellar, baryonic, and total mass dependence on the luminosity with the evolutionary stage of the galaxy. The luminosities of star-forming blue-cloud galaxies are dominated by young, massive stars. In contrast, quiescent, red-sequence galaxies have fewer massive stars and their luminosities are dominated by older, less massive stars. Thus, for a given luminosity, we observe blue-cloud galaxies to have a lower stellar mass than green-valley or red-sequence galaxies \citep{Carroll17}. 

The more evolved green-valley and red-sequence spirals resemble the elliptical galaxies more closely than the blue-cloud spirals, which are at an earlier stage of their evolution and have not yet converted their gas into stars \citep{Guo16}. Once the gas contribution is accounted for, there is a smaller difference in the baryonic mass. 

\begin{deluxetable}{lCCCCCC}
    \tablecaption{Parameters of the linear fits ($\log M = a M_r + b$) from Fig.~\ref{fig:mtot_mstar_mr_all}.
    \label{tab:fit_masses_Mr}}
    \tablehead{ & \multicolumn{2}{c}{$\log M_*$} & \multicolumn{2}{c}{$\log$\Mbar} & \multicolumn{2}{c}{$\log$\Mtot} \\
    \colhead{CMD classification} & \colhead{$a_*$} & \colhead{$b_*$} & \colhead{$a_{\rm bar}$} & \colhead{$b_{\rm bar}$} & \colhead{$a_{\rm tot}$} & \colhead{$b_{\rm tot}$}
    }
    \startdata
        Blue cloud spirals 
        & -0.444 \pm 0.004 & 1.369 \pm 0.083 
        & -0.403 \pm 0.005 & 2.291 \pm 0.089
        &  -0.332 \pm 0.005 & 4.137 \pm 0.089 \\
        Green valley spirals 
        & -0.427  \pm 0.012 & 1.861 \pm 0.249 
        & -0.424 \pm 0.014 & 1.986 \pm 0.271
        &  -0.386 \pm 0.013 & 3.133 \pm 0.265 \\
        Red sequence spirals 
        & -0.374 \pm 0.020 & 2.973 \pm 0.404 
        & -0.380 \pm 0.026 & 2.884 \pm 0.533
        &  -0.388 \pm 0.003 & 3.088 \pm 0.065\\
        Ellipticals 
        & -0.383 \pm 0.005 & 2.888 \pm 0.115 
        & \text{--} & \text{--}
        & -0.504 \pm 0.011 & 0.691 \pm 0.229 \\
    \enddata
\end{deluxetable}

\section{The Extended Baryonic Tully-Fisher Relation}\label{sec:btfr}

In order to construct the baryonic Tully-Fisher relation, we require baryonic masses and rotational velocities for the galaxies in the sample. As described in Sec.~\ref{sec:bar_mass_components}, we consider the stellar mass and the gas mass comprised of \HI, \Htwo, and He to be the dominant baryonic mass components of the spiral galaxies. For the baryonic mass of the ellipticals, we only consider the stellar mass, as we find their gas component to be typically $\lesssim 0.1M_*$. 

For spiral galaxies, dominated by rotational motion, we use $V(R_{90})$ for the rotational velocity of MaNGA galaxies, found as described in Sec.~\ref{sec:Spiral_fit}, and $V_{\rm eff}$ for the MaNGA elliptical galaxies. The evaluation of the effective rotational velocity for the elliptical galaxies is described in Sec.~\ref{sec:effectiveV}. For the TNG100 galaxies, we use the maximum rotational velocity in each galaxy's halo as the rotational velocity for the BTFR.

\subsection{The Extended BTFR for Matched MaNGA and TNG100 Galaxies}\label{sec:Fit_BTFR}

\begin{figure*}
    \centering
    \includegraphics[width=0.99\textwidth]{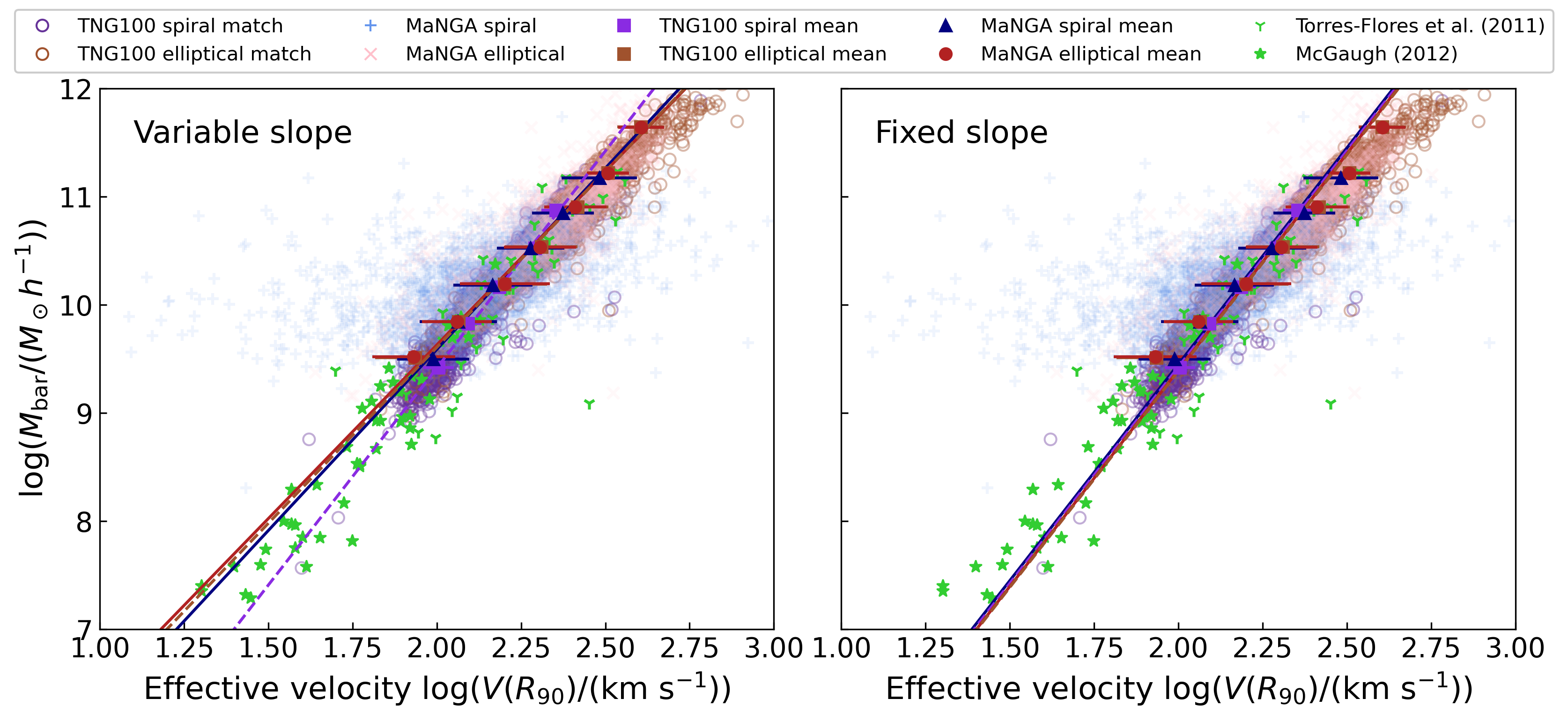}
    \caption{The BTFR for the MaNGA and matched TNG100 galaxies. In the linear fits the slope is fixed to 4 (right) and allowed to vary (left). Blue crosses are MaNGA spirals, pink x's are MaNGA ellipticals, purple circles are TNG100 spirals, and brown circles are TNG100 ellipticals. Navy triangles, red circles, purple squares, and brown squares are the means of fits to a Gaussian distribution for MaNGA spirals, MaNGA ellipticals, TNG100 spirals, and TNG100 ellipticals, respectively. The solid lines show fits to the MaNGA galaxies and dashed lines show fits to the TNG100 galaxies; see Table~\ref{tab:BTFR_fit_params} for the fit results.}
    \label{fig:BTFR_scatter}
\end{figure*}

The extended BTFR for the matched spiral and elliptical galaxies in the MaNGA and TNG100 catalogs is shown in Fig.~\ref{fig:BTFR_scatter}. Spiral galaxies exhibit more scatter compared to elliptical galaxies, presumably due either to uncertainties introduced by the inclusion of gas mass or the modeling of the rotational velocity for the former. We fit the BTFR of each sample by separating the galaxies into mass bins, fitting a Gaussian distribution to the data in each bin, whose mean and width are used as the $V_{\rm eff}$ points and their uncertainties. We then use Markov-Chain MC \citep{Goodman10} analysis to constrain the slope and $y$-intercept of the BTFR for both the spiral and elliptical populations. The results are shown in Fig.~\ref{fig:BTFR_scatter} (left) and summarized in Table~\ref{tab:BTFR_fit_params}. We find similar behavior between all four samples. The slopes of the $\Lambda$CDM-based simulation samples agree with the corresponding data samples.

\subsection{Consistency with $\Lambda$CDM and MOND}

In agreement with our observations, $\Lambda$CDM predicts a BTFR with a slope between 3 and 4 and allows for deviations from linear behavior. In Fig.~\ref{fig:BTFR_scatter} there is a hint of a turnover at the high end of the distribution, though it is not statistically significant. In contrast, MOND predicts a constant slope of exactly 4. To test the predictions of MOND, we fix the slope of the BTFR to 4, leaving the $y$-intercept free, and use the goodness of fit to evaluate the consistency of the data with this model.

The results of these fits are shown in Fig.~\ref{fig:BTFR_scatter} (right), and the parameters are given in Table~\ref{tab:BTFR_fit_params}. See Appendix~\ref{sec:btfr_mcmc} for the corresponding corner plots. When we fix the slope to 4, we find that the quality of the fit deteriorates for all samples except the TNG100 spirals. However, the difference in $\chi^2$ between the variable and fixed slope is not sufficient to exclude the MOND hypothesis.

In Fig.~\ref{fig:BTFR_scatter}, we show the eight BTFR fits. For comparison, we also include galaxies from two other BTFR studies, \cite{McGaugh:2012ac} and \cite{TorresFlores11}. Galaxies dominated by gas were selected for these studies, so their baryonic masses are lower than in the MaNGA sample. While the sample size is small, these galaxies follow our BTFR of the higher mass MaNGA and TNG100 samples.

\begin{figure*}
    \centering
    \includegraphics[width=0.99\textwidth]{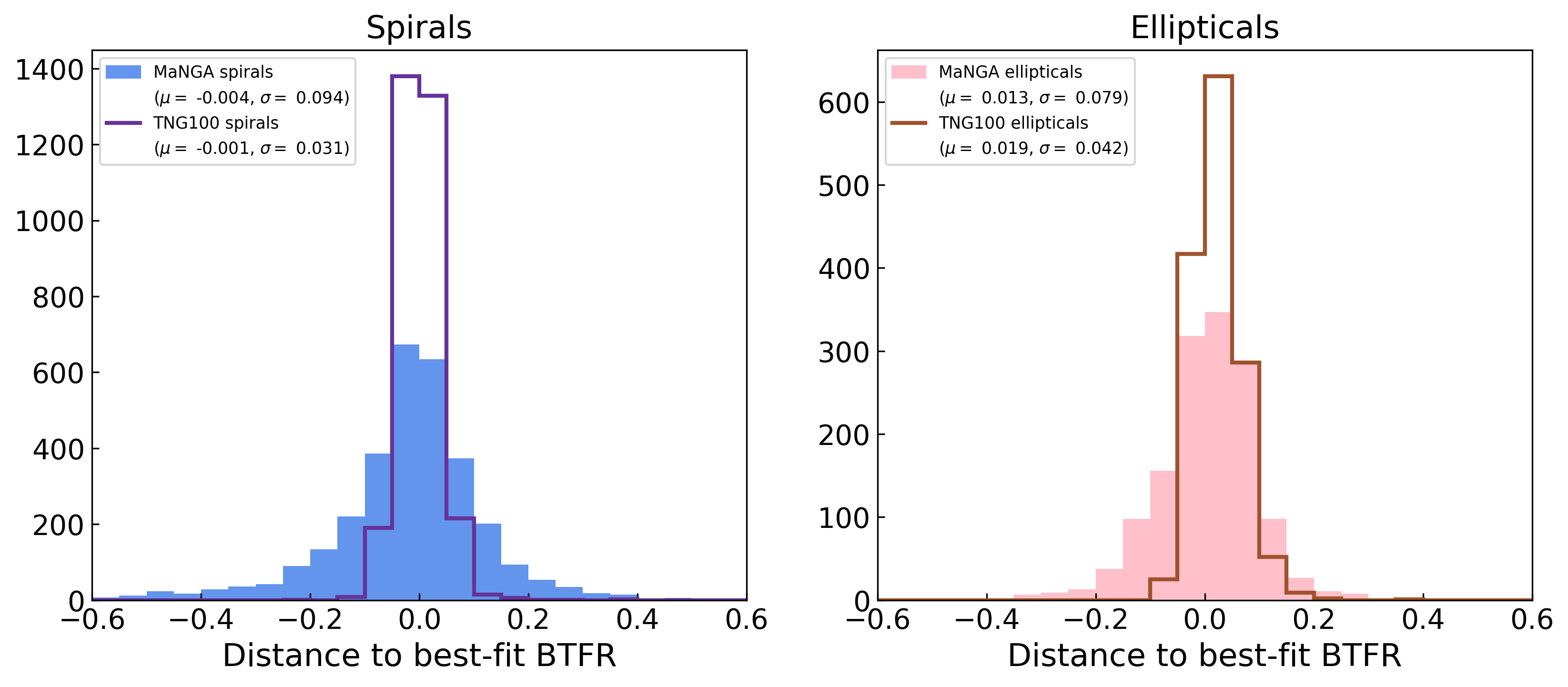}
    \caption{Perpendicular distance in the $\log(M_{\rm bar})$--$\log(V_{\rm eff})$ space to the best-fit BTFR with variable slope for spiral galaxies (left) and elliptical galaxies (right). Each histogram is fit to a Gaussian distribution with the mean and $\sigma$ given in the figure.}
    \label{fig:BTFR_residuals}
\end{figure*}

To quantify the spread in the BTFR, we calculate the perpendicular distance to the best-fit variable slope BTFR for each sample, shown in Fig.~\ref{fig:BTFR_residuals}. We fit a Gaussian distribution to the residual of each sample with $\sigma$ given in the figure caption. We define the difference in the spread in quadrature between the MaNGA and TNG100 samples as the spread introduced to the BTFR due to observational effects, $\sigma_{\rm obs}$. For the spirals, we find $\sigma_{\rm obs} = 0.090$ and for the ellipticals $\sigma_{\rm obs} = 0.067$. The scatter of the spirals is likely larger due to the additional uncertainties introduced by gas mass estimates and rotation curve fits that the elliptical galaxies do not suffer from.

\begin{deluxetable*}{lCCCc}
    \tablecaption{Parameters of the linear ($y=ax+b$) and quadratic ($y=cx^2+ax+b$) fits to the BTFR, where $y=\log M_{\rm bar}$, $x=\log V_{\rm rot}$ for spiral galaxies and TNG100 elliptical galaxies, and $x = \log V_{\rm eff}$ for the MaNGA elliptical galaxies, from Figs.~\ref{fig:BTFR_scatter}, \ref{fig:full_manga_btfr}, and \ref{fig:TNG_BTFR}.
    \label{tab:BTFR_fit_params}}
    \tablehead{
        \colhead{Sample} & \colhead{$a$} & \colhead{$b$} & \colhead{$c$} & \colhead{$\chi^2_\nu$}}
    \startdata
        TNG100 spirals & 4.01 \substack{+0.72 \\ -0.54} & 1.39 \substack{+1.17 \\ -1.57} & & 0.084\\
        MaNGA spirals & 3.35 \substack{+1.17 \\ -0.68} & 2.89 \substack{+1.54 \\ -2.62} & &  0.012\\
        TNG100 ellipticals & 3.25 \substack{+0.60 \\ -0.45} & 3.10 \substack{+1.08 \\ -1.44} & & 0.067\\
        MaNGA ellipticals & 3.21 \substack{+0.60 \\ -0.44} & 3.21 \substack{+1.04 \\ -1.45} & & 0.065\\
        \hline
        TNG100 spirals (fixed slope) & 4 & 1.42 \pm 0.08 & & 0.063\\
        MaNGA spirals (fixed slope) & 4 & 1.45 \pm 0.18 & & 0.090 \\
        TNG100 ellipticals (fixed slope) & 4 & 1.73 \pm 0.15 & & 0.347\\
        MaNGA ellipticals (fixed slope) & 4 & 1.40 \pm 0.15 & & 0.373\\
        \hline
        All MaNGA spirals & 3.55 \substack{+1.28 \\ -0.67} & 2.43 \substack{+1.55 \\ -2.59} & & 0.037\\
        All MaNGA ellipticals & 3.21 \substack{+0.72 \\ -0.49} & 3.18 \substack{+1.16 \\ -1.70} & & 0.087\\
        All MaNGA galaxies & 3.54 \substack{+0.65 \\ -0.48} & 2.46 \substack{+1.14 \\ -1.57} & & 0.104\\
        \hline
        All TNG100 $\log$(\Mbar) $> 9$ & 3.57 \substack{+0.48 \\ -0.37} & 2.36 \substack{+0.81 \\ -1.04} & & 0.051 \\
        All TNG100 & 9.37 $\pm$ 2.77 & -4.08 $\pm$ 2.86 & 
                    -1.29 $\pm$ 0.66 
                     & 0.148\\
    \enddata
\end{deluxetable*}

\begin{figure}
    \centering
    \includegraphics[width=0.49\textwidth]{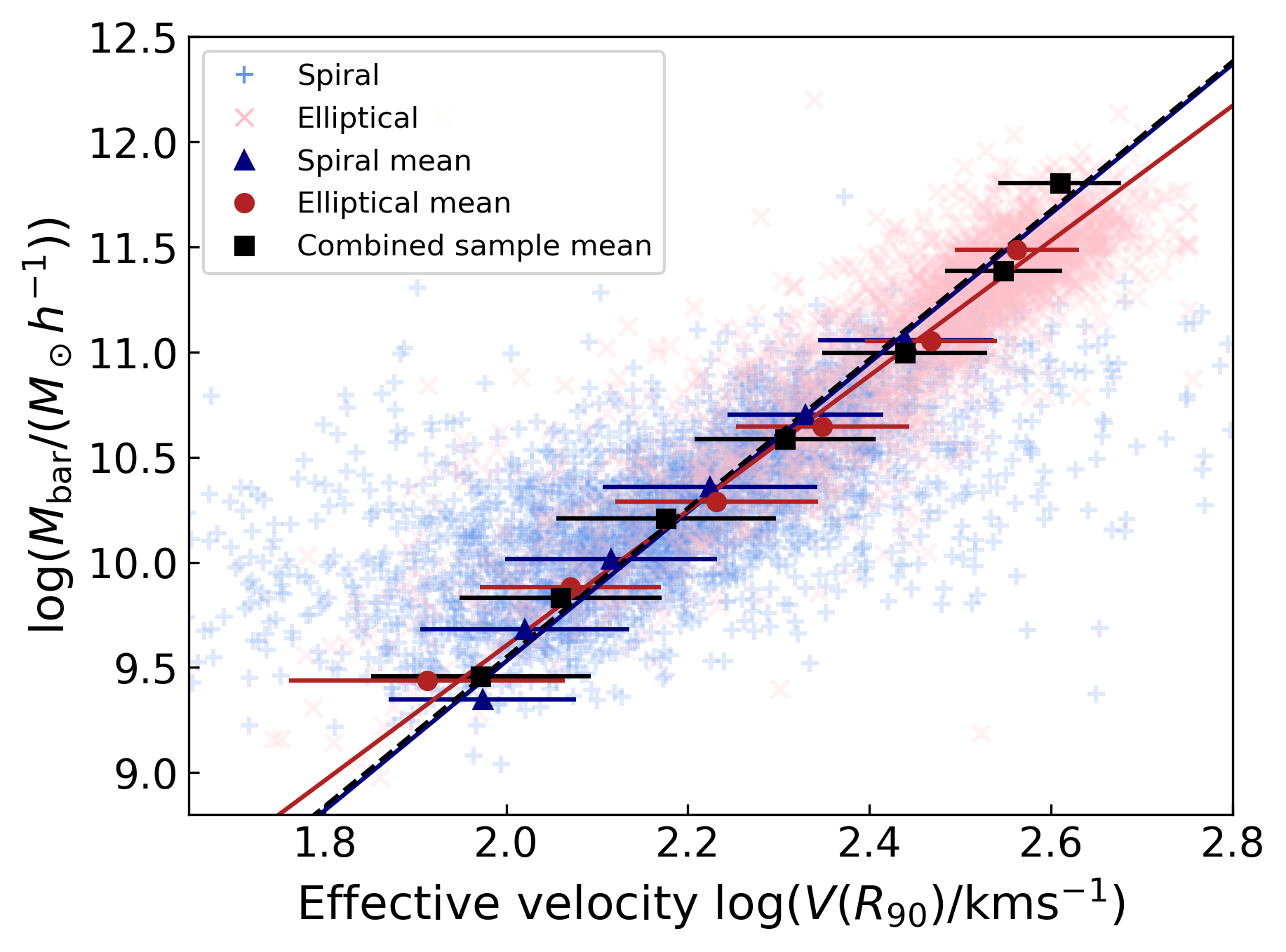}
    \caption{Joint BTFR for the MaNGA spiral and elliptical sample. Blue crosses are MaNGA spirals with \HI detections and pink x's are MaNGA ellipticals. The navy triangles (spirals), red circles (ellipticals), and black squares (combined sample) show the means from fits to a Gaussian distribution in each bin. The navy and red line show the linear fit to the spiral and elliptical galaxies and the black dashed line is the linear fit to the combined sample. See Table~\ref{tab:BTFR_fit_params} for the fit results.}
    \label{fig:full_manga_btfr}
\end{figure}

\subsection{Joint BTFR of the MaNGA Galaxies}

In Fig.~\ref{fig:full_manga_btfr}, we show the BTFR for the full MaNGA sample. This includes all MaNGA spirals with \HI masses and all MaNGA ellipticals (see middle row of Table~\ref{tab:sample_cuts}). We fit for the the slope and $y$-intercept of the BTFR for the spiral and elliptical samples separately and find agreement between the two BTFR fits, presented in Table~\ref{tab:BTFR_fit_params}. We also find that the BTFR fits to the full MaNGA sample agree with the BTFR fits for the TNG100 matched subsamples, indicating that these subsamples are representative of the overall population.

We also construct a joint BTFR for the full MaNGA sample, combining the spiral and elliptical samples together. See Appendix~\ref{sec:btfr_mcmc} for the corresponding corner plots. We find a slope of $3.54 \substack{+0.65\\ -0.48}$ which is consistent with both the prediction from $\Lambda$CDM (a slope between 3 and 4) and MOND (a slope of 4). The uncertainty on the fit is too large to distinguish between the two models.

\subsection{BTFR of the Full TNG100 Sample}\label{sec:Fit_BTFR_TNG100-1}

\begin{figure}
    \centering
    \includegraphics[width=0.49\textwidth]{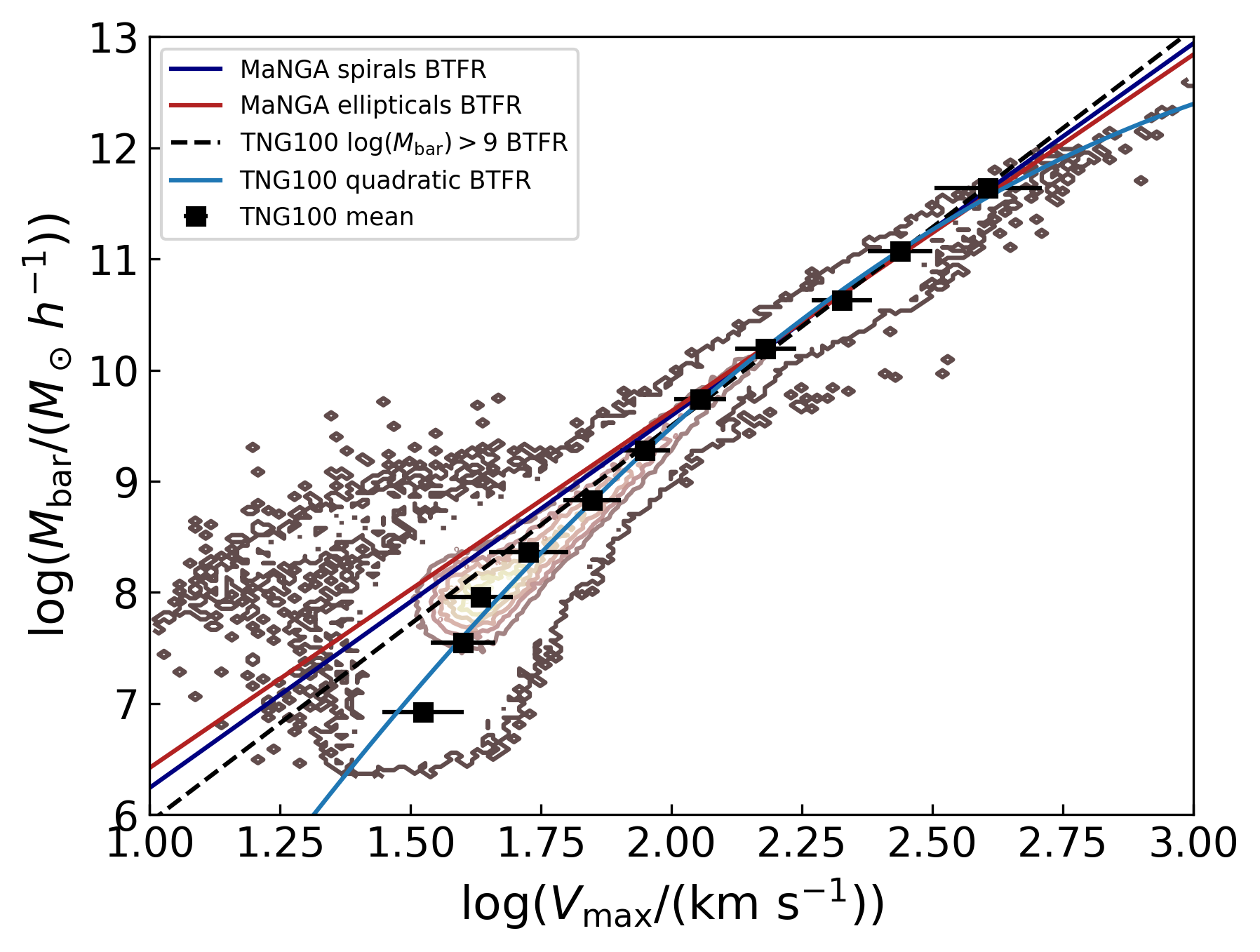}
    \caption{The BTFR for all TNG100 galaxies. The contours show the TNG100 galaxies, with the black squares as the population means. The blue and red solid lines are the fits from Fig.~\ref{fig:BTFR_scatter}. The black dashed line is a linear fit to the TNG100 galaxies with $\log(M_{\rm bar}) > 9$, and the light blue line is a quadratic fit for the full mass range. See Table~\ref{tab:BTFR_fit_params} for the fit results.}
    \label{fig:TNG_BTFR}
\end{figure}

The target selection of the MaNGA survey results in the sample containing galaxies with $\log(M_{\rm bar}) \gtrsim 9$, where the behavior of the BTFR is linear. To understand the $\Lambda$CDM prediction for lower mass galaxies, in Fig.~\ref{fig:TNG_BTFR} we show the BTFR for all galaxies in the TNG100 simulation, regardless of whether they have a match in MaNGA sample. The fit parameters for the slope and $y$-intercept of the BTFR for galaxies with $\log M_{\rm bar}>9$ are given in Table~\ref{tab:BTFR_fit_params}, and the corresponding corner plot is given in Appendix~\ref{sec:btfr_mcmc}.

The population of low mass galaxies deviate significantly from the linear behavior, predicting rotational velocities that are too high for their baryonic masses given a linear BTFR, potentially due to their large dark matter content. To capture this behavior, we perform a quadratic fit \citep{Neill14, Kourkchi20} across the full mass range, with parameters given in Table~\ref{tab:BTFR_fit_params}. This non-linear behavior contrasts with that observed in the samples from \cite{McGaugh:2012ac} and \cite{TorresFlores11}, which fall roughly along the same slope as the matched MaNGA and TNG100 samples. Therefore, in order to conclusively distinguish between the predictions of MOND and $\Lambda$CDM, future analyses must include a significant number of dwarf galaxies to fill in this low \Mbar region.

\section{Conclusions} \label{sec:conclusion}

We use the kinematic data of elliptical and spiral galaxies observed in SDSS MaNGA DR17 to study the effect of the galaxy evolutionary stage on the relation between stellar, baryonic, and total masses and luminosity, and to construct the baryonic Tully-Fisher relation. We estimate the stellar mass by modeling it as the sum of an exponential sphere and exponential thin disk. We find that, for a given $r$-band absolute magnitude, elliptical galaxies have larger stellar masses than spiral galaxies. When the spiral galaxies are separated by their CMD classification, we find that the discrepancy decreases as the spiral galaxy population evolves from the blue cloud to the red sequence. The similarity between the red-sequence spirals and the ellipticals supports the theory that elliptical galaxies are the end-stage of galaxy evolution. We also find that, after including the mass contribution from gas, the slope of the baryonic mass--luminosity relation for blue-cloud spirals matches that for the green-valley and red-sequence spirals more closely, demonstrating that this evolution is largely a conversion of gas into stars.

We model the rotation of spiral galaxies using H$\alpha$ kinematics data. From the models, we extract the rotational velocity at $R_{90}$ for each spiral galaxy as the rotational velocity to be used in the BTFR. We use the stellar velocity dispersion of the elliptical galaxies to estimate their total masses using the virial theorem. Using these total masses, we estimate each elliptical galaxy's effective rotational velocity at $R_{90}$, which we define as the rotational velocity a spiral galaxy of the same mass would have.

To verify this methodology we match the observed MaNGA galaxies to galaxies in the IllustrisTNG 100-1 simulation by finding the closest galaxy in the $u-r$ color--stellar mass plane. We find good agreement between the total mass of MaNGA galaxies to the dynamical mass within $R_{90}$ for the corresponding TNG100 galaxies, indicating that the virial theorem, evaluated using the half-light radius, only provides an estimate of the total mass within $R_{90}$. The effective rotational velocity of elliptical MaNGA galaxies are also found to be in good agreement with the maximum rotational velocity in the halo of the matched TNG100 galaxies. 

Introducing the effective rotational velocity for the elliptical galaxies allowed us to construct an extended BTFR for both elliptical and spiral MaNGA galaxies. This significantly increases the sample size and extends the range of probed baryonic masses by about 1.5~dex. The baryonic mass of the gas-poor elliptical galaxies is dominated by stellar mass, and as a result suffers from fewer uncertainties compared to the baryonic mass of the spiral galaxies, where different gas components must be evaluated in a separate measurement or parameterized in separate samples. When comparing the BTFR of the MaNGA galaxies to that of the matching TNG100 galaxies, we find a good agreement for the elliptical and spiral galaxies. For the MaNGA spiral galaxies, we find a large scatter in the relation compared with the other samples, likely due to uncertainties in the estimation of gas mass and the rotation curve fits.

$\Lambda$CDM allows for a variation in the slope of the BTFR between 3 and 4, deviations from linear behavior, and some scatter due to the variable contribution from dark matter. In contrast, MOND, which offers an alternative to dark matter theory, predicts a BTFR slope of exactly 4, with no deviation from linear behavior and no significant scatter. We observe a significant scatter for the spiral galaxies, which could be explained by uncertainties in the gas estimation, and a very tight BTFR for the elliptical galaxies. We test the agreement of data with both models by fitting it to a linear function with a variable ($\Lambda$CDM) and fixed to 4 (MOND) slope. The former provides a very good fit to data, and there is a hint of a turnover at the high-mass end observed, but it is not statistically significant. We find that while $\Lambda$CDM agrees better with the data, MOND cannot be excluded within the mass range of MaNGA galaxies. 

We construct an extended BTFR for the combined sample of MaNGA spiral and elliptical galaxies. We also construct a BTFR for all galaxies in the TNG100 simulation and find strong agreement between the MaNGA and TNG100 BTFRs. Both samples result in a slope that is consistent with MOND at $\log (M_{\rm bar}) > 9$. However, the TNG100 simulation shows a low-mass population of galaxies, not surveyed in MaNGA, that deviates significantly from the linear BTFR. Further observations in this mass range are necessary to test $\Lambda$CDM and MOND predictions of the BTFR.

\section*{Acknowledgements}

N.R. acknowledges support from the Feinberg Research Award through the Department of Physics \& Astronomy at the University of Rochester. We acknowledge support from the Department of Energy under the grant DE-SC0008475.

This publication makes use of the GBT Legacy Archive. The National Radio Astronomy Observatory and Green Bank Observatory are facilities of the U.S. National Science Foundation operated under cooperative agreement by Associated Universities, Inc.

This project makes use of the MaNGA-Pipe3D data products. We thank the IA-UNAM MaNGA team for creating this catalogue, and the Conacyt Project CB-285080 for supporting them.

Funding for the Sloan Digital Sky Survey IV has been provided by the Alfred P. Sloan Foundation, the U.S. Department of Energy Office of Science, and the Participating Institutions. SDSS-IV acknowledges support and resources from the Center for High-Performance Computing at the University of Utah. The SDSS web site is www.sdss.org.

SDSS-IV is managed by the Astrophysical Research Consortium for the Participating Institutions of the SDSS Collaboration including the Brazilian Participation Group, the Carnegie Institution for Science, Carnegie Mellon University, the Chilean Participation Group, the French Participation Group, Harvard-Smithsonian Center for Astrophysics, Instituto de Astrof\'isica de Canarias, The Johns Hopkins University, Kavli Institute for the Physics and Mathematics of the Universe (IPMU) / University of Tokyo, the Korean Participation Group, Lawrence Berkeley National Laboratory, Leibniz Institut f\"ur Astrophysik Potsdam (AIP), Max-Planck-Institut f\"ur Astronomie (MPIA Heidelberg), Max-Planck Institut f\"ur Astrophysik (MPA Garching), Max-Planck-Institut f\"ur Extraterrestrische Physik (MPE), National Astronomical Observatories of China, New Mexico State University, New York University, University of Notre Dame, Observat\'ario Nacional / MCTI, The Ohio State University, Pennsylvania State University, Shanghai Astronomical Observatory, United Kingdom Participation Group, Universidad Nacional Aut\'onoma de M\'exico, University of Arizona, University of Colorado Boulder, University of Oxford, University of Portsmouth, University of Utah, University of Virginia, University of Washington, University of Wisconsin, Vanderbilt University, and Yale University.

The IllustrisTNG simulations were undertaken with compute time awarded by the Gauss Centre for Supercomputing (GCS) under GCS Large-Scale Projects GCS-ILLU and GCS-DWAR on the GCS share of the supercomputer Hazel Hen at the High Performance Computing Center Stuttgart (HLRS), as well as on the machines of the Max Planck Computing and Data Facility (MPCDF) in Garching, Germany.

\software{
Astropy \citep{astropy:2013, astropy:2018, astropy:2022}, 
Corner \citep{corner}, 
emcee \citep{emcee},
Matplotlib \citep{matplotlib},
NumPy \citep{numpy}, 
SciPy \citep{scipy}
}


\appendix 
\section{Fitting the Baryonic Tully-Fisher Relation}\label{sec:btfr_mcmc}

\begin{figure}[h]
    \centering
    \includegraphics[width=0.32\linewidth]{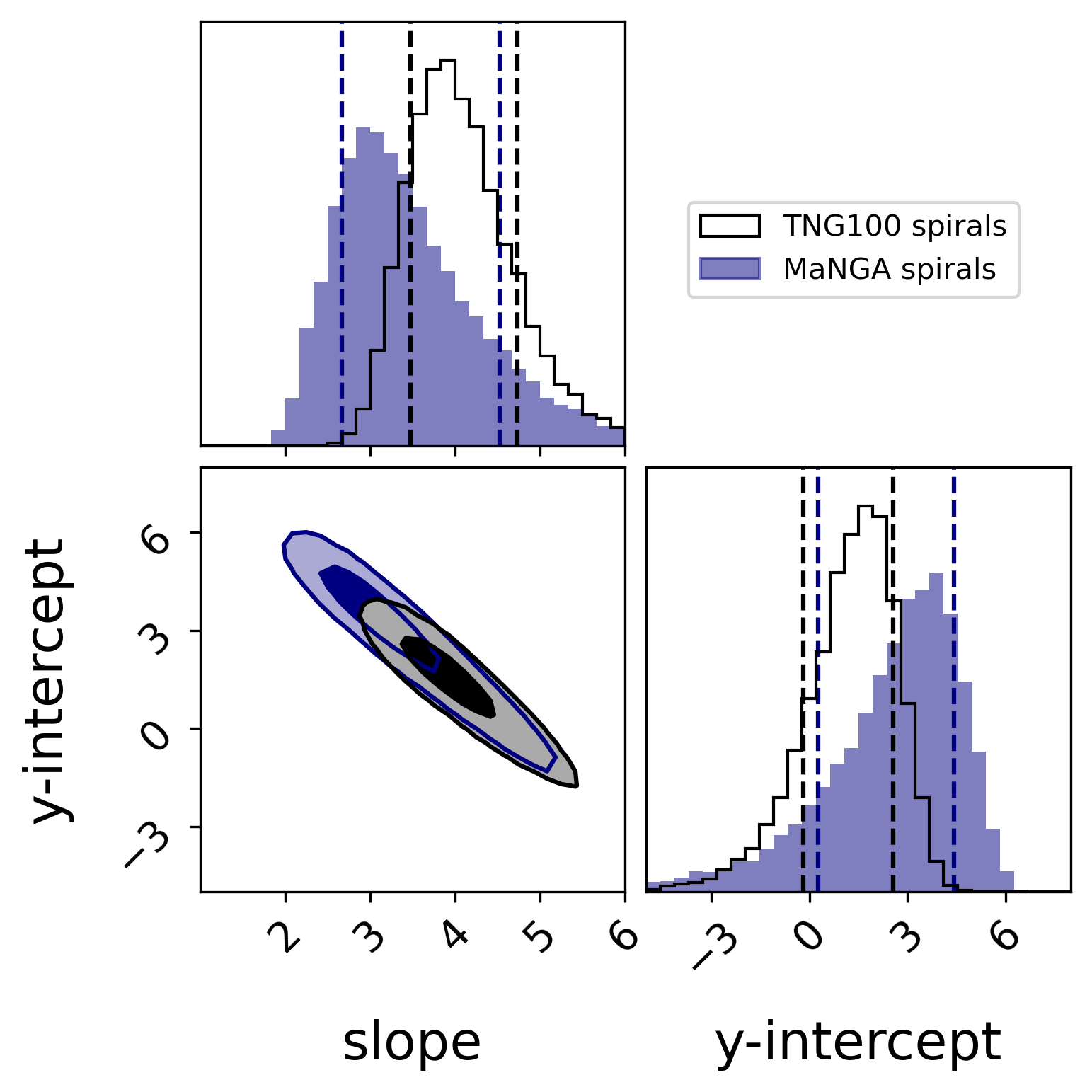}
    \includegraphics[width=0.32\linewidth]{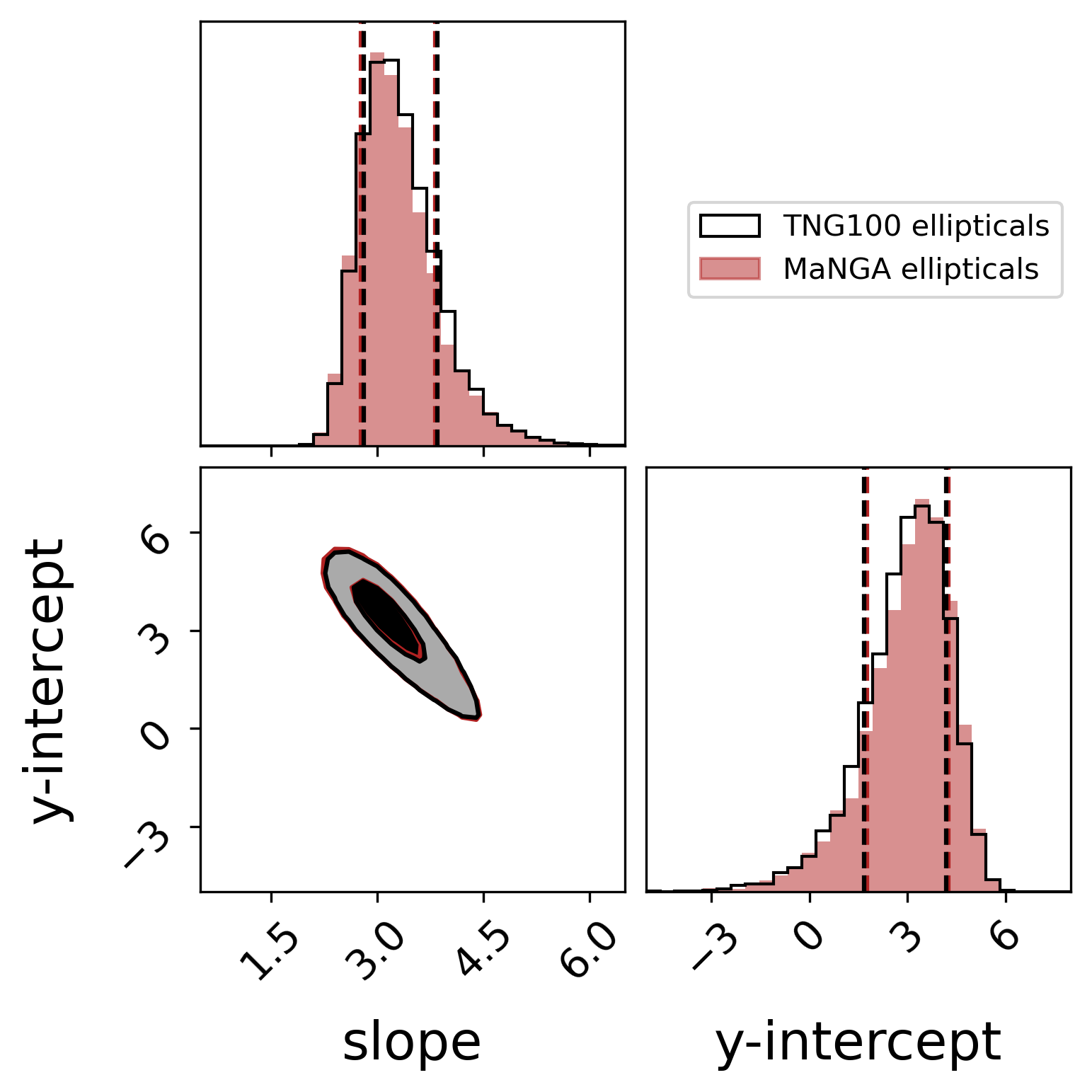}
    \includegraphics[width=0.32\linewidth]{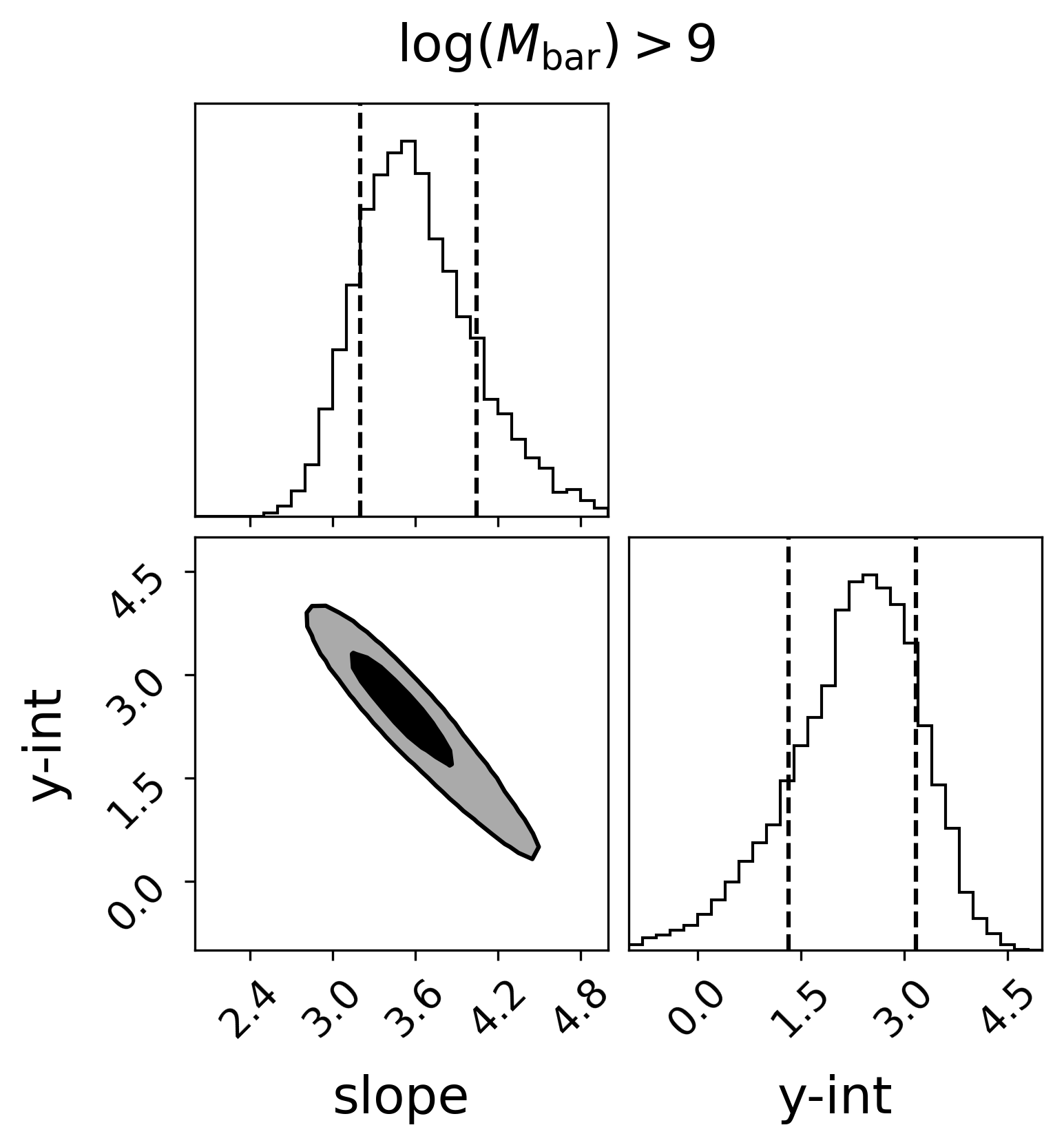}
    \caption{Corner plots for the BTFR fits for the MaNGA and TNG100 matched spiral (left) and elliptical (center) samples, and the full TNG100 simulation with $\log$\Mbar $> 9$. The dashed lines show the 16th and 84th quantiles for each parameter.}
    \label{fig:manga_sim_corner}
\end{figure}

\begin{figure}[h]
    \centering
   \includegraphics[width=0.45\linewidth]{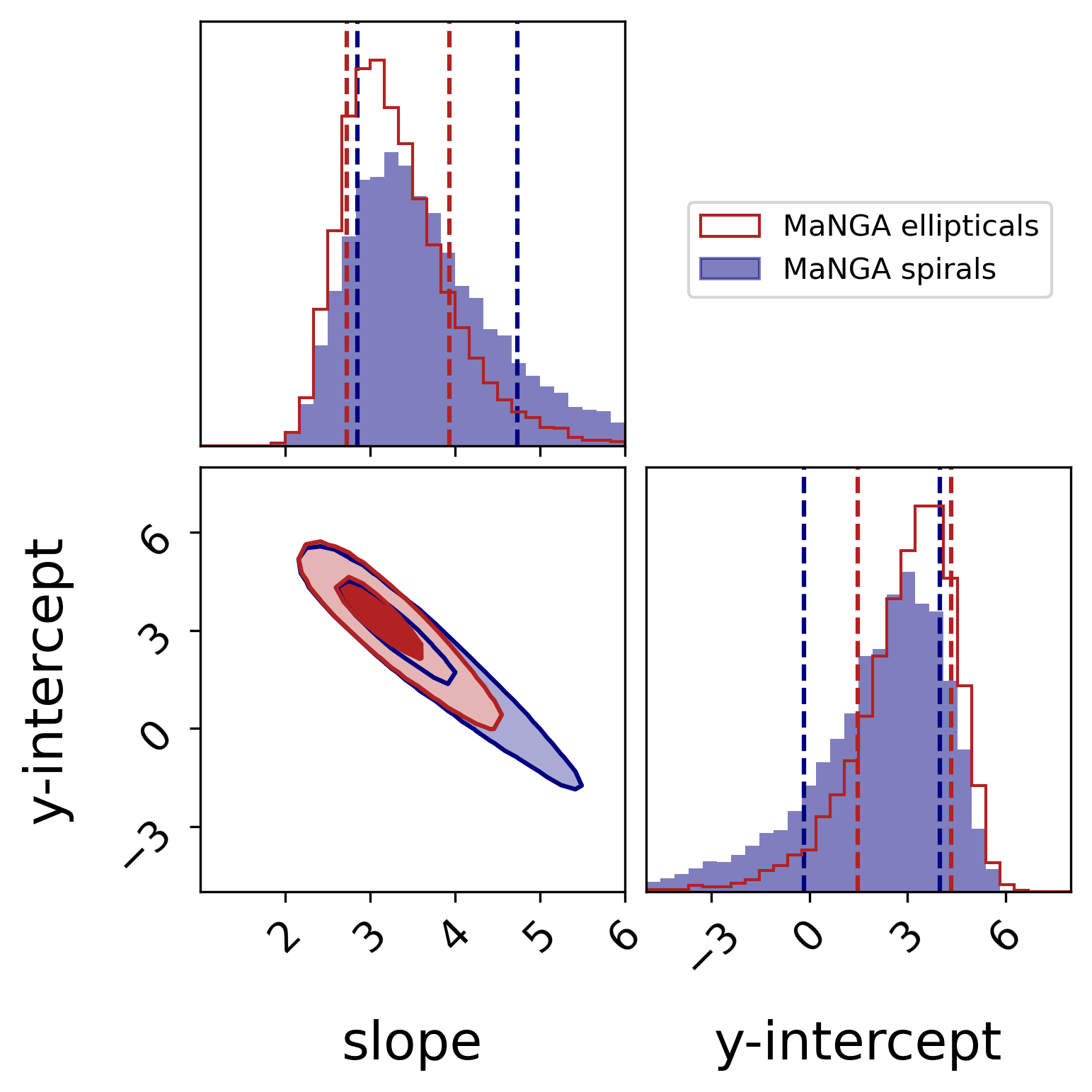}
    \includegraphics[width=0.45\linewidth]{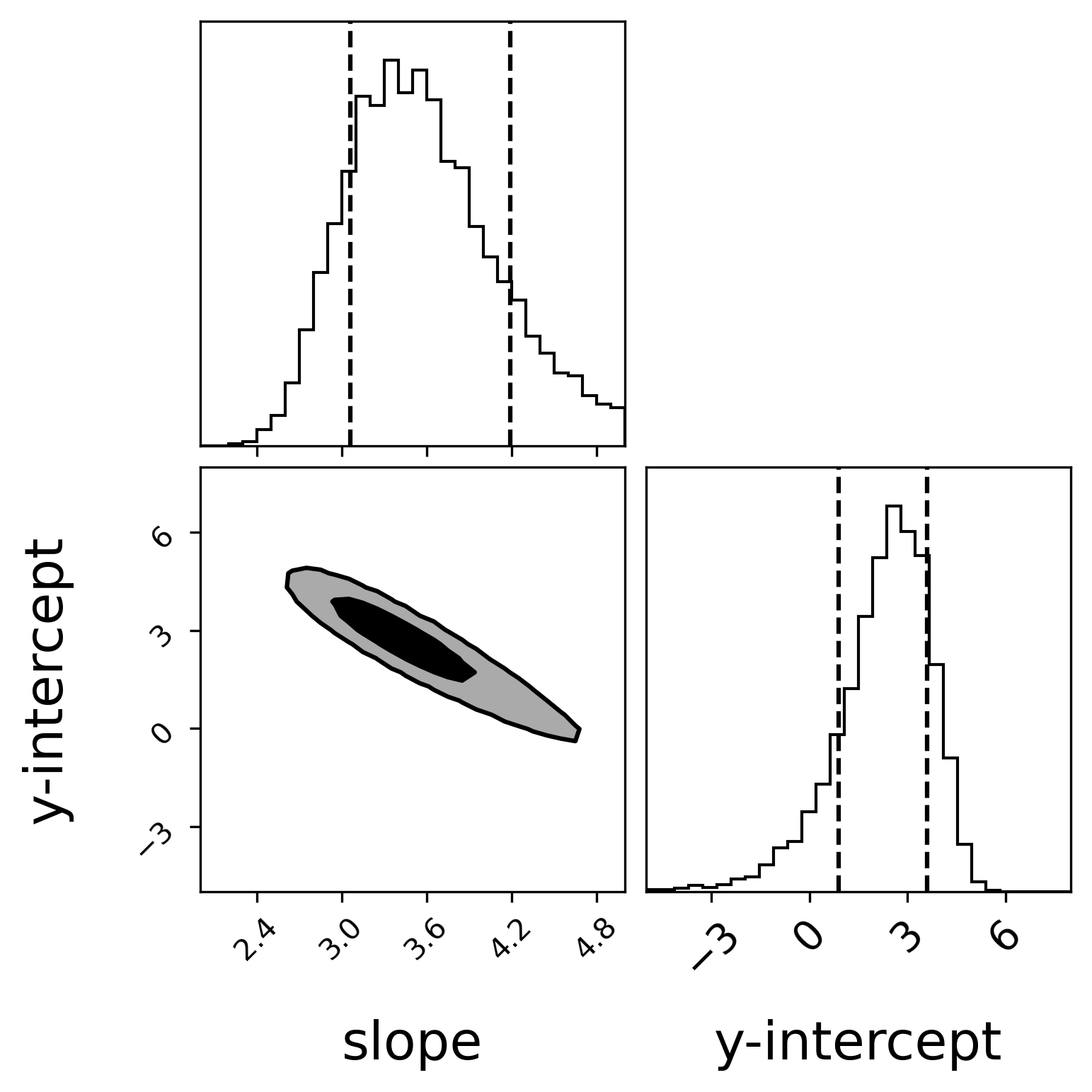}
    \caption{Corner plots for the BTFR fits for the full MaNGA spiral and elliptical samples (left) and the joint fit to the combined sample (right). The dashed lines show the 16th and 84th quantiles for each parameter.}
    \label{fig:manga_all_corner}
\end{figure}

We use an MCMC algorithm to find the slope and $y$-intercept of the baryonic Tully-Fisher relation for subset of MaNGA spirals and ellipticals matched to TNG100 and their corresponding galaxies in TNG100. The results of the fits are shown in the corner plots in Fig.~\ref{fig:manga_sim_corner}, with the values given in Table~\ref{tab:BTFR_fit_params}. The right panel of Fig.~\ref{fig:manga_sim_corner} shows the results of the BTFR fit for all TNG100 galaxies with $\log M_{\rm bar} > 9$. Fig.~\ref{fig:manga_all_corner} shows the results of the fits for the BTFR of the full MaNGA sample for ellipticals and spirals and the extended BTFR for the combined sample.

\bibliographystyle{aasjournalv7}
\bibliography{Ravi1225_sources}

\end{document}